%% file: main.tex
\documentclass{article}

\usepackage[english]{babel}
\usepackage[utf8x]{inputenc}
\usepackage[T1]{fontenc}
\usepackage{natbib}
\usepackage{cancel}
\usepackage{soul}
 \bibpunct{(}{)}{;}{a}{,}{,}

\usepackage{fullpage}

\usepackage{amsmath,amsthm, amssymb}
\usepackage{graphicx}
\usepackage{hyperref}

\usepackage{algorithm,algpseudocode}
\usepackage{multirow}

\usepackage{setspace}

\usepackage{caption}
\usepackage{subcaption}
\usepackage{multicol}

\newtheorem{theorem}{Theorem}
\newtheorem{corollary}{Corollary}[theorem]

\newtheorem{proposition}{Proposition}

\theoremstyle{definition}
\newtheorem{definition}{Definition}
\newtheorem{assumption}{Assumption}
\newtheorem{condition}{Condition}

\newtheorem{example}{Example}

\newcommand\independent{\protect\mathpalette{\protect\independenT}{\perp}}
\def\independenT#1#2{\mathrel{\rlap{$#1#2$}\mkern2mu{#1#2}}}

\def\-{\text{-}}
\def\Bern{\text{Bern}}

\def\revisionadd{}

\ifdefined\revisionadd

\else

\fi

\ifdefined\revision

\else

\fi

\ifdefined\revisioncut
  \newcommand{\edits}[1]{\textcolor{red}{\st{#1}}}
\else
  \newcommand{\edits}[1]{}
\fi

\title{Flexible sensitivity analysis for observational studies without observable implications
\protect\thanks{Alexander M. Franks is an Assistant Professor of
  Statistics at the University of California, Santa Barbara
  (\href{mailto:afranks@pstat.ucsb.edu}{afranks@pstat.ucsb.edu}).
  Alexander D'Amour is now a Research Scientist at Google AI,
  Cambridge,  MA (\href{mailto:alexdamour@google.com}{alexdamour@google.com}); most of this work wascompleted when he was a Neyman Visiting Assistant Professor of Statistics at the University of California, Berkeley.  Avi Feller is an Assistant Professor of Public Policy and Statistics at the University of California, Berkeley (\href{mailto:afeller@berkeley.edu}{afeller@berkeley.edu}).
We thank Edo Airoldi, Carlos Cinelli, Peng Ding, Sharad Goel, Chad Hazlett, Jennifer Hill, Jared Murray, Sam Pimentel, Don Rubin, and Ravi Shroff as well as participants at ACIC 2018 for thoughtful comments and discussion. \newline\indent $^\dagger\,\,$These authors contributed equally to this work.}}

\author{Alex Franks$^\dagger$\\UCSB \and Alex D'Amour$^\dagger$\\UC Berkeley \and Avi Feller\\UC Berkeley}

\date{This draft: \today}

\begin{document}
\maketitle

\begin{centering}
\end{centering}

\thispagestyle{empty}
\pagenumbering{gobble}

\begin{abstract}
\input{sections/sec-abstract.tex}
\vspace{2em}

\end{abstract}

\clearpage
\pagenumbering{arabic}
\input{sections/sec-introduction.tex}
\input{sections/sec-extrap-causal.tex}
\input{sections/sec2-example-specs.tex}
\input{sections/sec-analysis.tex}


\input{sections/sec-discussion.tex}

\newpage
\appendix
\input{sections/appendix.tex}

\input{sections/app-figures.tex}

\clearpage
\singlespacing
\bibliographystyle{chicago}
\bibliography{refs,sensitivity_latent,refs_old_sens}

\end{document}

%% file: sections/sec-abstract.tex
A fundamental challenge in observational causal inference is that assumptions about unconfoundedness are not testable from data. 
Assessing sensitivity to such assumptions is therefore important in practice.
Unfortunately, some existing sensitivity analysis approaches inadvertently impose 
restrictions that are at odds with modern causal inference methods, which emphasize flexible models for observed data.
To address this issue, we propose a framework that allows (1) flexible models for the observed data and (2) clean separation of the identified and unidentified parts of the sensitivity model.
Our framework extends an approach from the missing data literature, known as Tukey's factorization, to the causal inference setting.
Under this factorization, we can represent the distributions of unobserved potential outcomes in terms of unidentified selection functions that posit an unidentified relationship between the treatment assignment indicator and unobserved potential outcomes.
The sensitivity parameters in this framework are easily interpreted, and we provide heuristics for calibrating these parameters against observable quantities.
We demonstrate the flexibility of this approach in two examples, where we estimate both average treatment effects and quantile treatment effects using Bayesian nonparametric models for the observed data. 

%% file: sections/sec-introduction.tex
\doublespacing

\section{Introduction}
\label{sec:intro}
Causal inference generally requires two distinct elements: modeling observed potential outcomes and making assumptions about missing potential outcomes. 
While researchers can investigate the first element with standard model-checking techniques, the data are uninformative about the second element, which can only be probed via sensitivity analysis.
In principle, a sensitivity analysis quantifies how results change under different assumptions about unobserved potential outcomes --- without affecting the observed data model \citep{linero2017bayesian, gustafson2018sensitivity}.
Unfortunately, such ``clean'' sensitivity analyses are often difficult to construct under common sensitivity analysis frameworks, sometimes leading to analyses where assumptions about the sensitivity analysis inadvertently impose constraints on the observed data model.
This tension puts sensitivity analysis at odds with modern causal inference approaches that incorporate flexible observed data models, such as recent approaches that feature Bayesian nonparametric regression or other machine learning methods  \citep{hill2012, athey2017, hahn2017bayesian}. 

In this paper, we propose a sensitivity analysis framework for observational studies that cleanly separates model checking from sensitivity analysis.
Specifically, we propose a factorization of the joint distribution of potential outcomes, known as \emph{Tukey's factorization}, that separates factors into those that are nonparametrically identified and those that are completely unidentified from the data. 
This factorization expresses the unidentified factors in terms of selection functions, which are easily interpreted, and a copula that characterizes the dependence between potential outcomes, which we can ignore for a broad class of common estimands. 
Taken together, we can then interpret the implied distribution on missing potential outcomes as an extrapolation from the observed data distribution.

Our main contribution is develop a practical workflow for this factorization in causal inference, thus extending Tukey's factorization, which has a long history in the missing data literature, to the observational study setting. 
First suggested by John Tukey \citep[recorded in][]{holland-notes}, variants of the approach have been known by a number of names including exponential tilting \citep{Birmingham2003,rotnitzky2001,Scharfstein1999}, non-parametric (just) identified (NPI) models \citep{robins2000sensitivity}, the extrapolation factorization \citep{linero2017bayesian} and Tukey's factorization \citep{franks2016non}. In this paper we use the term ``Tukey's factorization'' since our approach explicitly adapts the method proposed by \citet{franks2016non} to observational causal inference.  Tukey's factorization also bears close resemblance to missing data methods based on weighting, including importance weight methodology \citep{riddles2016propensity} and inverse probability weighting models for sensitivity analysis  \citep[e.g.][]{zhao2017sensitivity}.
While \citet{robins2000sensitivity} also proposed to apply Tukey's approach to unobserved confounding in observational studies, applications of their proposal appear to be limited to the context of clinical trials with dropout \citep{rotnitzky2001,scharfstein2003incorporating}.

%

The workflow we propose has a number of practical strengths for assessing sensitivity to unmeasured confounding.
First, the framework naturally accommodates a large class of commonly-applied observed data models, including models for complex data.
We demonstrate this flexibility by conducting sensitivity analysis with a range of modern statistical models including Bayesian Additive Regression Trees (BART) for response surface estimation \citep{dorie2016flexible} and Dirichlet processes mixture (DPM) models for flexible residual distributions.  We also demonstrate the use of the factorization with semi-continuous data using zero-inflated models.  
Second, the framework is computationally cheap.
Because the factorization separates observed data modeling from sensitivity parameters, an investigator only needs to fit the observed data model once, and can apply our sensitivity analysis \emph{post hoc}.
Finally, the sensitivity parameters defined in our framework are easily interpreted, facilitating model specification and parameter calibration.
In particular, the sensitivity parameters can be calibrated against variation explained by covariates in a standard propensity score model, which are already familiar to many investigators.



The paper proceeds as follows.
Section \ref{sec:setup} introduces the formal setup for sensitivity analysis and highlights key issues involving identifiability of sensitivity parameters.
Section \ref{sec:tukey_causal} introduces Tukey's factorization for causal inference and provides theoretical justification for applying the approach to a common set of estimands. 
Section \ref{sec:example_specs} defines a flexible, convenient model specification, the logistic-mixture exponential family model, and explores technical properties. 
Section \ref{sec:calibration} discusses heuristics for interpreting and calibrating the sensitivity parameters in this model.
Section \ref{sec:analysis} applies Tukey's factorization to two examples, demonstrating the flexibility of the approach on a range of estimands with several Bayesian nonparametric estimation approaches.
Finally, Section \ref{sec:discussion} discusses open issues and possible extensions. The appendix contains technical details and some additional results from the applied analyses.

\section{Setup and Overview of Sensitivity Analysis} 
\label{sec:setup}


\subsection{Setup and Notation}

We describe our approach using the potential outcomes framework \citep{neyman1923, rubin1974}. For outcome $Y_i \in \mathcal{Y}$ for unit $i$ and binary treatment, let $Y_i(0)$ and $Y_i(1)$ denote that unit's potential outcomes if assigned to control or treatment, respectively. Let $T_i$ denote a binary treatment indicator and $X_i$ denote observed covariates. For compactness, we often write $Y_i(t),\: t \in \{0, 1\}$ to denote the outcome for treatment level $t$. Assuming SUTVA \citep{rubin1980comment}, we can then write the observed outcome as
$
Y_i^{\text{obs}} = T_i Y_i(1) + (1-T_i) Y_i(0)
$.
Finally, we write the propensity score as
$
e(x) = P(T_i = 1 \mid X_i = x).
$
We assume an infinite population of iid units, from which we sample triplets $([Y_i(0), Y_i(1)], T_i, X_i)$ and observe triplets $(Y^{\text{obs}}_i, T_i, X_i)$. Given this, we suppress the subscript $i$ unless otherwise noted.

We focus on two classes of population estimands.
First, we consider average treatment effects on the whole population as well as on the treated and control populations:
\begin{align*}
\tau^{ATE} &:= E[Y(1) - Y(0)] = E[Y(1)] - E[Y(0)]\\
\tau^{ATT} &:= E[Y(1) - Y(0) \mid T = 1] = E[Y(1) \mid T = 1] - E[Y(0) \mid T = 1]\\
\tau^{ATC} &:= E[Y(1) - Y(0) \mid T = 0] = E[Y(1) \mid T = 0] - E[Y(0) \mid T = 0].
\end{align*}
Second, we consider quantile treatment effects,  $$\tau_q =  Q_q(Y(1)) - Q_q(Y(0))$$
\noindent the difference in the $q$-th treatment quantile, $Q_q(Y(1))$, and control quantile, $Q_q(Y(0))$.

Each of these estimands is a contrast between the marginal complete-data outcome distributions $f(Y(1))$ and $f(Y(0))$.
For each $t$, the complete-data distribution for each potential outcome can be written as a mixture of the distribution of observed and missing outcomes:
$$
f(Y(t) \mid X) = f(T = t \mid X)f^{\rm obs}_t(Y(t) \mid T = t, X) + f(T = 1-t \mid X)f^{\rm mis}_t(Y(t) \mid T = 1-t, X).
$$

All factors in this expression are identified except for $f^{\rm mis}_t$, which is completely uninformed by the data.
Identifying these estimands thus requires untestable assumptions that characterize $f^{\rm mis}_t$.
Sensitivity analysis probes the robustness of estimates to these assumptions.
Specifically, we consider sensitivity analyses for observational studies where the investigator wishes to test robustness to violations of the unconfoundedness assumption.
\begin{assumption}[Unconfoundedness]
\label{assn:unconfounded}
$[Y(0), Y(1)] \independent T \mid X $.
\end{assumption}
\noindent Unconfoundedness implies that $f^{\rm obs}_t = f^{\rm mis}_t$, and is thus sufficient to identify our estimands of interest.  Broadly, sensitivity analysis for violations of unconfoundedness proceeds by parameterizing the conditional dependence between partially-observable potential outcomes $[Y(0), Y(1)]$ and $T$ given covariates $X$.
The parameters of this dependence are called \emph{sensitivity parameters}.
Investigators can then report how causal effect estimates change when the sensitivity parameters are allowed to vary within a selected range of plausible values.
The plausibility of the specific values of sensitivity parameters is ultimately determined externally to the data analysis, e.g., by domain expertise.

\subsection{Summary of Approach}
\label{sec:summary}


We propose a method for model-based sensitivity analysis that explicitly factorizes the joint distribution of $([Y(0), Y(1)], T \mid X)$ in terms of the observed outcome distributions $f^{\rm obs}_t$.
Specifically, we parameterize the missing outcome distributions as a tilt of the observed outcome distribution
\begin{align}
f^{\rm mis}_{\psi, t}(Y(t) \mid T=1\-t, X) \propto \frac{f_\psi(T=t\mid Y(t), X)}{f_\psi(T=1\-t\mid Y(t), X)}f^{\rm obs}_t(Y(t) \mid T=t, X).
\label{eq:fmis extrapolation}
\end{align}
We explore key features of this factorizaton in subsequent sections.
This specification defines the complete-data distributions $f_\psi(Y(t))$ implicitly.
The fraction can be interpreted as importance weights that transform the observed outcome distribution $f^{\rm obs}_t$ into the missing outcome distribution $f^{\rm mis}_{\psi, t}$ parameterized by $\psi$. 
Notably, the observed data distributions in this approach, $f_t^{\rm obs}$, are free of the sensitivity parameter $\psi$, which implies a clean separation of model checking and sensitivity analysis.
%
In practice, sensitivity analysis with Equation \eqref{eq:fmis extrapolation} involves a number of non-trivial implementation details,
including imposing constraints on normalizing constants and setting and interpreting sensitivity parameters. 

\subsection{Related Methods}
\label{sec:lit review}
There is an extensive literature on assessing sensitivity to departures from unconfoundedness, dating back at least to the foundational work of \citet{cornfield1959smoking} on the link between smoking and lung cancer.
The approach we take in this paper fits broadly within the \emph{model-based} sensitivity analysis framework, largely building on \citet{rosenbaum1983assessing}.
Examples of alternative sensitivity analysis approaches include 
\citet{rosenbaum2009amplification},~\citet{diaz2013sensitivity},~\citet{ding2016sensitivity}, and \citet{zhao2017sensitivity}.

Our approach most closely follows recent proposals to enable model-based sensitivity analysis with flexible outcome models, especially \citet{dorie2016flexible}, who use Bayesian Additive Regression Trees (BART) models for flexible outcome modeling \citep[see also][]{carnegie2016assessing}.
We contrast their approach with ours in Section \ref{sec:analysis_nhanes}.
Similarly, \citet{jung2018algorithmic} combine flexible, black-box modeling with sensitivity in the setting of algorithmic decision making.
These methods extend the so-called \emph{latent confounder approach} of sensitivity analysis \citep{rosenbaum1983assessing} by specifying a parametric but flexible latent variable model for the complete data.
The latent confounder approach is highly intuitive, and in cases where these is strong scientific prior knowledge, may be preferable to the approach we propose here. For example, when the parameters in the model are scientifically meaningful, this formulation may be necessary to elicit priors for these parameters.
As we discuss next, however, such models can introduce issues around identified sensitivity parameters. 
While some modern approaches incorporate parameterizations that mitigate these identification concerns,\footnote{For example, the model in \citet{jung2018algorithmic} has sensitivity parameters that depend only weakly on the observed data because they focus exclusively on the setting with a binary outcome and employ sensitivity parameters that vary smoothly across covariates.}
the degree to which sensitivity parameters are decoupled from the observed data model can be difficult to assess in practice.

A number of previous methods have also sought to separate model checking from sensitivity analysis in the context of missing data.
Specifically, our approach is most similar to non-parametric (just) identified (NPI) models proposed by \citet{robins2000sensitivity}, so called because they put no constraints on the observed data model, but identify a complete-data distribution when the sensitivity parameters are fixed.
NPI models were proposed using a variant of Tukey's factorization, but have been applied primarily to longitudinal missing data problems and clinical trials with dropout; see, e.g., \citet{rotnitzky2001} and \citet{scharfstein2003incorporating}.
\citet{robins2000sensitivity} suggested that NPI models could be applied to observational studies with unobserved confounding, but to the best of our knowledge, did not pursue this method in applications.
Our paper builds a practical sensitivity analysis method based on this theoretical suggestion, in part by proposing a specification that reduces some of the technical hurdles in the original \citet{robins2000sensitivity} proposal; see Section \ref{sec:tukey_causal}. 

Our method is also related to a line of Bayesian missing data methods summarized by \citet{linero2017bayesian}.
These methods specify a Bayesian model for observed data, and explore a set of ``identifying restrictions'' that map the observed data model to complete data models.
Tukey's factorization in \eqref{eq:fmis extrapolation} is an example of an identifying restriction, although the authors generally choose alternative restrictions in their applications.
Similar to the NPI literature, these methods do not appear to have been applied to observational studies, with their causal applications restricted to clinical trials with dropout \citep{linero2015flexible,linero2017biometrika}.  Finally, in a Bayesian approach that aligns well with ours, \citet{hahn2016bayesian} propose flexible, nonparametric models that separate the observed data model from sensitivity parameters, also in the context of missing data.

\subsection{Separating Sensitivity Analysis from Model Checking}
\label{sec:separation}

A prime motivation for our sensitivity analysis framework is to separate sensitivity analysis from model checking, that is, from assessing the fit of different modeling assumptions to the observed data. 
In practice, many popular approaches to sensitivity analysis blur the line between sensitivity analysis and model checking by introducing sensitivity parameters that are informed by the observed data.
Sensitivity analyses of this type can be difficult to interpret, however. Such methods typically relax the unconfoundedness assumption by introducing a parametric specification, which can in turn fundamentally change the relationship between confounding and identification.
This can also introduce new types of sensitivity that go unexamined by the analysis itself, resulting in conclusions whose credibility depend on strong prior knowledge about the parametric specification.
We demonstrate some of these pathologies in the context of a latent confounder model in the following example.



\begin{figure}	

	\centering
		\includegraphics[width=0.5\textwidth]{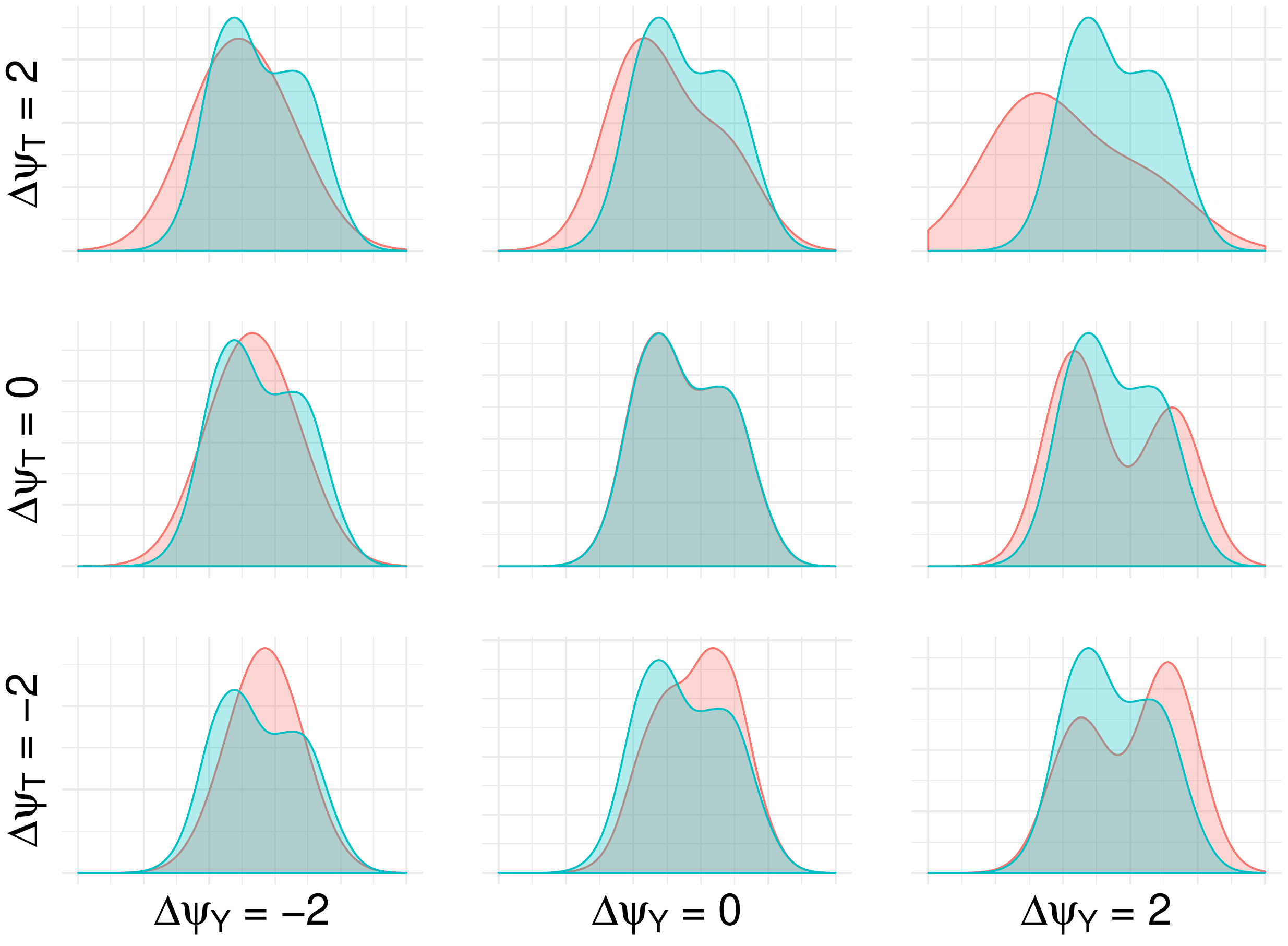}
	\caption{ Illustrations of sensitivity parameter identification in Example 1, via variation in posterior predictive distributions at different sensitivity parameter settings. We plot the true (blue) and inferred (red) observed control outcome densities, $f(Y(0) \mid T=0)$, implied by the normal outcome-binary latent confounder model with sensitivity parameters $\psi_T$ and $\psi_Y$. $\Delta \psi_T$ and $\Delta \psi_y$ represent the difference between the true and assumed sensitivity parameters. The observed data densities are only correctly inferred when we assume the true values of the sensitivity parameters $(i.e. \Delta \psi_T = \Delta \psi_Y = 0$). Note that the true potential outcome distributions $Y(0)$ are a mixture of normal distributions with the same component means and different variances.
}
\label{fig:misfit}
\end{figure}

\begin{example}[Normal Outcome, Binary Confounder]
\label{ex:normal outcome binary confounder}
Suppose we have a study with a continuous outcome and no covariates.
We make the assumption that treatment was randomly assigned according to a Bernoulli design, but it is plausible that there exists a latent class that confounds the study.  To test the robustness of our conclusions to the presence of such a latent class, we propose a sensitivity analysis by introducing a binary latent confounder.  The model is parameterized as follows: 
\begin{align*}
U &\sim \Bern(\xi_u)\\
T \mid U &\sim \text{Bern}(g(\alpha + \psi_T U))\\
Y(t) \mid U &\sim N(\mu_{t} + \psi_Y U, \sigma^2).
\end{align*}
When $(\psi_T, \psi_{Y}) = (0, 0)$, the model reduces to random assignment to treatment. 
Importantly, the observed data distribution depends on the sensitivity parameters $\psi$.
To see this, let $h_{\psi_T} := f(U = 1 \mid T)$.
The distribution of observed outcomes is a two-component mixture of normals for $t = \{0,1\}$:
\begin{align}
\label{eqn:dorie_model}
Y(t) \mid T=t &\sim h_{\psi_T}N(\mu_{t} + \psi_Y, \sigma^2) + (1-h_{\psi_T})N(\mu_{t}, \sigma^2)
\end{align}
\noindent The observed and missing potential outcome distributions, $f(Y(t) \mid T = t, X)$ and $f(Y(t) \mid T = 1-t, X)$, respectively, have the same mixture components but different mixture weights, where $\psi_Y$ determines the difference between the component means.
Importantly, the mixture weights, $h_{\psi_T(t)}$ and component means are identifiable under relatively weak assumptions \citep[see][]{everitt1985mixture}. 

In Figure~\ref{fig:misfit}, we demonstrate a range of data fits obtained from this model.\footnote{In this simulation we set $\xi_u = 1/2, \alpha=0, \mu_t=0, \psi_T=-1, \psi_Y=5$ and $\sigma^2=1$.}
As is common practice, we fit the model multiple times for a range of sensitivity parameters and plot the true (blue) and inferred (red) observed potential outcome densities for the control potential outcomes. 
Most settings of the sensitivity parameters result in severe misfit of the observed data, and would likely be rejected by a competent analyst if they were considered the ``main'' model for the observed data.
In this case, the sensitivity analysis operates as a model checking exercise more than an exploration of relaxed identification assumptions. 
In fact, based on this parametric specification, the investigator could reject the hypothesis that the study satisfies unconfoundedness (Assumption~\ref{assn:unconfounded}) based on model fit.
In the case where the investigator had strong scientific basis to believe the parametric form of the latent confounder model, this would be useful information. However, in the absence of a priori knowledge about this specification, the unconfoundedness assumption is not tested in full generality.

More elaborate forms of this model appear throughout the literature, including extensions that incorporate covariates and nonparametric regression models \citep{imbens2003sensitivity,dorie2016flexible}.
Under these specifications, the residuals of the observed outcomes $Y(t) - E[Y(t) \mid X, T = t]$ are distributed as a mixture of normals, and the identification issue noted in this example remains.
\end{example}

There are several additional examples of parametric models inducing testable implications of otherwise untestable assumptions in the missing data and causal literature.
For example, \citet{littlebook} note that ignorability can be tested in the Heckman selection model  \citep{heckman1979} as a result of its Gaussian parameterization.
Similarly, \citet{linero2017bayesian} describe a detailed case, due to \citet{Kenward_Selection_1998}, where slight changes in the tail thickness of a parametric complete-data model specification result in different test-based conclusions about ignorability.

In addition to complicating interpretation, mixing model checking with sensitivity analysis also has practical implications.
First, sensitivity analysis with identified sensitivity parameters is computationally expensive because each setting of the sensitivity parameters requires that the model be re-fit \citep[see][for discussion]{hahn2016bayesian}.
This is particularly onerous when a sensitivity analysis strategy is employed with modern nonparametric strategies that are relatively expensive to fit.
Second, exploring a range of distinct model fits in a \emph{post hoc} sensitivity analysis raises the spectre of data snooping.
In a valid sensitivity analysis, the investigator can tune an observed data model in a held-out sample.
By contrast, in a sensitivity analysis with identified sensitivity parameters, the investigator can instead choose the model fit that is the most favorable to their conclusions, and declare it to be the ``main'' model whose robustness is being checked  \citep[][p. 172, ``What Is Not a Sensitivity Analysis?'']{Rosenbaum_Observation_2017}.
Finally, in the context of Bayesian sensitivity analysis, \citet{gustafson2018sensitivity} note that identified sensitivity parameters introduce a tension between the prior information used to calibrate the sensitivity parameters and the information from the observed data that make posterior uncertainties difficult to interpret.
For example, in the context of a clinical trial with dropout, \citet{scharfstein2003incorporating} consider and reject a parametric complete-data model with an identified sensitivity parameter because, under this model, the observed data fit implies implausibly large selection bias.

%% file: sections/sec-extrap-causal.tex
\section{Tukey's Factorization for Causal Inference}
\label{sec:tukey_causal}

We now turn to applying Tukey's factorization to assessing sensitivity to unobserved confounding, paying special attention to the structural differences between this new setting and the more common missing data setting. We show that these structural differences imply a distinct Tukey factorization for causal inference, which includes a copula that characterizes the dependence between potential outcomes. Importantly, we show that there is a broad class of estimands, including the ATE and QTE, whose sensitivity does not depend on this copula. For these estimands, sensitivity analysis can proceed by simply treating the observational study as though it were two separate missing data problems. In short, this result can be summarized as: ``causal inference is missing data twice''.

\subsection{Tukey's Factorization with One Potential Outcome}
\label{sec:tukey_one_PO}

We begin by demonstrating Tukey's factorization on only one arm of an observational study.
Specifically, we examine the joint distribution of one potential outcome $Y(t)$ and the treatment indicator $T$. 
This case is analogous to a missing data problem, where Tukey's factorization has been applied previously \citep{Birmingham2003, Scharfstein1999, franks2016non, linero2017bayesian}.
Tukey's factorization of this joint distribution is
\begin{equation}
\label{tukeysFactorization}
f_\psi(Y(t), T \mid X) = f^{\rm obs}(Y(t)\mid T=t, X)f(T=t \mid X)\cdot \frac{f_\psi(T\mid Y(t), X)}{f_\psi(T=t\mid Y(t), X)}.
\end{equation}
Here, the first two factors constitute the observed data density, which is nonparametrically identified, while the final factor is determined by the \emph{selection function}, which is unidentified but easily interpreted. 
In our approach, we parameterize the selection function with sensitivity parameters $\psi$.
%
%
Thus, the unidentified selection function fully determines the relationship between the observed outcome distribution and the distribution of missing outcomes.

The validity of this factorization requires two technical conditions. First, following \citet{robins2000sensitivity}, we need to impose some mild restrictions on the selection factors.


\begin{condition}[Integral constraints] 
\label{cond:int_contraint}
For each $t$, the normalizing constant in \eqref{eq:fmis extrapolation} satisfies:
\begin{align}
\label{eqn:selection_constraint}
\int_{\mathcal Y} f^{\rm obs}(Y(t)\mid T=t, X)\cdot \frac{f_\psi(T=1\-t\mid Y(t), X)}{f_\psi(T=t\mid Y(t), X)} dY(t) = \frac{f(T= 1\-t \mid X)}{f(T=t \mid X)}.
\end{align}
\end{condition}

\noindent This condition arises by integrating both sides of \eqref{tukeysFactorization} with respect to $Y(t)$. This condition restricts the class of selection functions that can be used within our framework. While this restriction has lead to technical challenges explored in earlier work \citep[see][for additional discussion]{robins2000sensitivity}, the parameterizations we propose in Section \ref{sec:logit-ef} satisfy this constraint automatically.

%
%


Second, we need an outcome overlap condition, which ensures that the missing data distribution can be expressed as an extrapolation of the observed data distribution.
\begin{condition}[Outcome Overlap Condition]
\label{cond:missing support}
The support of the missing potential outcomes is a subset of the support of the observed potential outcomes.
That is,
$$
P(Y(t) \in A \mid T = 1-t, X) > 0 \Rightarrow P(Y(t) \in A \mid T = t, X) > 0
$$
for all sets $A$ in the outcome sample space $\mathcal Y$.
\end{condition}
\noindent 
As opposed to Condition~\ref{cond:int_contraint}, this condition is an assumption about the true data-generating process and cannot be enforced by model specification.
We discuss this assumption further in Section \ref{sec:discussion}.




\subsection{Tukey's Factorization with Both Potential Outcomes}

\subsubsection{General factorization}

We now extend Tukey's factorization to the observational study setting.  In particular, we show that the joint distribution of potential outcomes $(Y(0), Y(1))$ and treatment $T$ can be uniquely specified by supplementing the distribution of the observed data with three unidentified models: a model for treatment given $Y(1)$ alone; a model for treatment given $Y(0)$ alone; and a copula that specifies the dependence between $Y(0$) and $Y(1)$ given $T$.


Under Conditions~\ref{cond:int_contraint} and~\ref{cond:missing support}, the joint density $f(T, [Y(0), Y(1)] \mid X)$ can be decomposed into two univariate complete-data densities and a copula.
The derivation follows as a consequence of applying Tukey's factorization to the marginal densities $f(T, Y(0) \mid X)$ and $f(T, Y(1) \mid X)$:
\begin{align}
\label{eqn:extrapolation_causal}
f(T, [Y(0), Y(1)] \mid X) &= 
f(Y(0) \mid T, X)\cdot f(Y(1) \mid T, X) \cdot f(T \mid X) \cdot \frac{f(Y(0), Y(1) \mid T, X)}{f(Y(0) \mid T, X) f(Y(1) \mid T, X)} \nonumber \\
&= \frac{f(Y(0), T \mid X)}{f(T\mid X)} \cdot \frac{f(Y(1), T \mid X)}{f(T \mid X)} \cdot
\frac{f(Y(0), Y(1) \mid T, X)}{f(Y(0) \mid T, X) f(Y(1) \mid T, X)} \cdot f(T\mid X) \nonumber \\
&= \begin{aligned}[t]
  &f(Y(0)\mid T=0, X)f(T=0 \mid X)\cdot \frac{f(T \mid Y(0), X)}{f(T=0\mid Y(0), X)} \cdot\\
  & f(Y(1)\mid T=1, X)f(T=1 \mid X)\cdot \frac{f(T\mid Y(1), X)}{f(T=1\mid Y(1), X)}\cdot\frac{1}{f(T \mid X)} \cdot\\
  &c(F(Y(0)\mid T, X), F(Y(1)\mid T, X) \mid T, X)
\end{aligned}
\end{align}
\noindent where we use $F$ to represent a cumulative distribution function. We now define several important terms. First, we define the 
\emph{marginal selection factors} $f(T \mid Y(1), X)$ and $f(T \mid  Y(0), X)$, which specify the non-ignorable selection mechanism in each arm. Second, we define the \emph{conditional copula}
$$c(F(Y(0)\mid T, X), F(Y(1)\mid T, X) \mid T, X) = \frac{f(Y(0), Y(1) \mid T, X)}{f(Y(0) \mid T, X)f(Y(1) \mid T, X)}$$
as the copula density that characterizes the residual dependence between potential outcomes conditional on the assigned treatment. Apart from the marginal selection factors and the conditional copula, all other terms in \eqref{eqn:extrapolation_causal} are identifiable.


Equation \eqref{eqn:extrapolation_causal} implies that the density of the missing potential outcomes, conditional on the observed potential outcomes is 
%
\begin{align}
\label{eqn:miss_causal}
    f^{\rm mis}(Y(t) \mid T=(1\-t), Y(1\-t), X]) \;\;
&\propto \;\; f^{\rm obs}(Y(t)\mid T=t, X) \cdot \frac{f(T=(1\-t) \mid Y(t), X)}{f(T=t\mid Y(t), X)} \cdot \nonumber\\
&\qquad\quad c(F(Y(0)\mid T, X), F(Y(1)\mid T, X) \mid T, X).
\end{align}
We can then use this factorization to estimate the complete joint distribution, $f([T, Y(0), Y(1)] \mid X)$, and, in turn, estimate causal estimands of interest.

\subsubsection{Marginal Contrast Estimands}


So far, we have described the full set of unidentified factors necessary to specify the joint density of $f(T, [Y(0), Y(1)] \mid X)$.
We now show that for a broad set of estimands, one need not specify the conditional copula to construct a well-defined sensitivity analysis.
These results give us license to conduct sensitivity analysis for such estimands as though the observational study were two independent missing data problems.
%
We refer to the general class of estimands for which this invariance holds as \emph{marginal contrasts}.
%
 \begin{definition}
 \label{def:marginal contrast}
A causal estimand $\tau$ is a \emph{marginal contrast estimand} iff it can be completely characterized as a function of the marginal distributions of the potential outcomes $f(Y(0) \mid X)$ and $f(Y(1) \mid X)$.
Formally, letting $\tau$ be a functional of the joint distribution $f(T, [Y(0), Y(1)] \mid X)$, with slight abuse of notation, a marginal contrast satisfies
$
\tau(f(T, [Y(0), Y(1)] \mid X)) = \tau(f(Y(0) \mid X), f(Y(1) \mid X)).
$
 \end{definition}
 
This class includes common estimands include the Average Treatment Effect, $E[Y(1)] - E[Y(0)]$, and the Quantile Treatment Effect, $\tau_q =  Q_q(Y(1)) - Q_q(Y(0))$. This class, however, excludes estimands that depend on the joint distribution of potential outcomes, such as the proportion of units with positive treatment, the variance of the treatment effect, and unit-specific treatments \citep[see, for example,][]{heckman1997making,ding2018decomposing}.

 
%
\begin{theorem}
\label{thm:copula invariance}
Suppose the joint distribution $f(T, [Y(0), Y(1)] \mid X)$ admits Tukey's factorization in \eqref{eqn:extrapolation_causal}.
Further, suppose that $\tau$ is a marginal contrast estimand.
Then $\tau$ is uniquely defined by the marginal selection factors $f(T \mid Y(1), X)$ and $f(T \mid Y(0), X)$.
\begin{proof}
See appendix.
\end{proof}
\end{theorem}
%
\noindent Stated differently, marginal contrast estimands are invariant to the conditional copula $c(F(Y(0) \mid T, X), F(Y(1) \mid T, X) \mid T, X)$.  In likelihood-based estimation, the invariance of the estimand translates to invariance in estimation. In the special case of Bayesian inference, the following corollary establishes when the posterior distribution for marginal contrast estimands exhibits invariance to the specification of the copula.
This invariance holds under a distinct parameters condition that is common in Bayesian formulations of ignorability \citep{gelman2013bayesian}.
\begin{corollary}
\label{cor:bayesian}
In the setting of Theorem~\ref{thm:copula invariance}, suppose, in addition, that the parameters of the conditional copula $c(F(Y(0 \mid T, X)), F(Y(1)\mid T, X) \mid T, X)$ are distinct, i.e., \emph{a priori} independent, of the parameters of all other factors in \eqref{eqn:extrapolation_causal}.
Then the posterior distributions for marginal contrast estimands are invariant to the specification of the conditional copula in the model likelihood. 
\end{corollary}
%
%
%
 %
This establishes that the selection factors in each treatment arm are the \emph{only} unidentifiable factors that need to be specified when estimating marginal contrast estimands by Bayesian inference. 
Obviating the need for the copula is useful in practice because the dependence between potential outcomes is unobservable, even in experiments, and thus assumptions about this dependence are difficult to calibrate.
However, one would need to wrestle with this complication for some of the more general sets of estimands mentioned above (i.e. unit-specific treatment effects).  These estimands, which are \emph{not} marginal contrasts, arise for instance when estimating optimal treatment regimes \citep{klausch2018estimating}.  
We leave sensitivity analysis for estimands of this type to future work. 

%% file: sections/sec2-example-specs.tex
\section{Logistic Selection with Mixtures of Exponential Family Models}
\label{sec:example_specs}
\label{sec:logit-ef}

We now propose a simple but widely applicable class of models 
that addresses several practical challenges with implementing Tukey's factorization for causal inference. 
Specifically, we focus on \emph{logistic selection with mixtures of exponential families} (logistic-mEF models).
In these models, the marginal selection functions in each arm are specified as logistic in the potential outcomes, and the observed data is modeled with a mixture of exponential family distributions. 
Building on previous work, we show that this class of models automatically satisfies Condition~\ref{cond:int_contraint} and yields missing potential outcome distributions that are analytically tractable.
Importantly, logistic-mEF models include non-parametric observed data models like Dirichlet process mixtures (DPM) and BART.
Thus, beyond being analytically convenient, the results in this section directly apply to sensitivity analysis with many flexible modeling tools already used in causal inference.  We start by describing the setting in which the observed data densities belong to a single exponential family distribution, and then demonstrate how our formulation extends to mixtures of exponential families. We leave it to future work to explore specifications outside of the logistic-mEF family.

\subsection{Logistic selection specifications}

Under a logistic selection specification, we posit that the log-odds of receiving treatment are linear in some sufficient statistics of the potential outcomes, $s(Y(t))$:
\begin{align}
\label{eqn:logistic selection}
f(T = 1 \mid Y(t), X) &= \text{logit}^{-1}\hbox{}\{\alpha_t(X) + \gamma_t' s_t(Y(t))\},
\end{align}
where $\text{logit}^{-1}\hbox{}(x) = (1 + \exp(-x))^{-1}$.
This specification has sensitivity parameters $\gamma = (\gamma_0, \gamma_1)$, which describe how treatment assignment depends marginally on each potential outcome, and a parameter $\alpha_t(X)$ in each arm that is identified by the observed data once $\gamma_t$ is specified (discussed below).
Logistic selection is commonly used in latent confounder approaches to model the probability of treatment given the unobserved confounder.
Here, we take a similar approach but instead assume that the treatment probabilities are logistic in (sufficient statistics of) the potential outcomes.
Throughout this paper we assume the sensitivity parameters are independent of covariates. We return to this point in the discussion.

Beyond their interpretability, logistic selection specifications have several desirable technical properties, some of which have been explored previously in different settings.
Here, we adapt these results to our setting and notation.
First, for any specification of $\gamma_t$ that implies a proper missing data distribution, Condition~\ref{cond:int_contraint} is automatically satisfied. 
In particular, \citet{robins2000sensitivity} shows that in each arm $t$, for each $\gamma_t$, there exists a unique $\alpha_t(X)$ that satisfies Condition~\ref{cond:int_contraint}.
Second, \citet{rotnitzky2001} and \citet{scharfstein2003incorporating} show that the missing data distribution implied by a logistic selection specification in each arm $t$ is free of $\alpha_t(X)$.
Thus, having specified $\gamma_t$, it is not necessary to solve for $\alpha_t(X)$ explicitly.
Specifically,
\begin{align}
f^{\rm{mis}}(Y(t) \mid T = 1-t, X) = f^{\rm{obs}}(Y(t) \mid T = t, X)
    \frac{\exp(\gamma_t' s_t(Y(t)))}{C(\gamma_t, X)},\label{eq:logistic missing}
\end{align}
where the normalizing constant $C(\gamma_t, X) = \int_{\mathcal Y} \exp(\gamma_t's_t(Y(t))) f^{\rm{obs}}(Y(t) \mid T = t, X) dY(t)$.
Finally, in the observational causal inference setting, because Condition~\ref{cond:int_contraint} is satisfied independently for valid values of $\gamma_t$ in each arm, the sensitivity parameter vectors $\gamma_0$ and $\gamma_1$ are variation independent.

The primary practical difficulty in applying the logistic selection specification is computing the normalizing constant $C(\gamma_t, X)$ in \eqref{eq:logistic missing} for each specification of $\gamma_t$.
The normalizing constant is necessary for computing causal effects at each value of $\gamma_t$, as well as calibrating $\gamma_t$, as we discussion Section~\ref{sec:calibration}.
Dealing with this normalizing constant in practice can require either computationally expensive calculation or strong restrictions on the observed data model.
For example, \citet{scharfstein2003incorporating} address the normalizing constant by either modeling the observed data with its empirical CDF, for which the integration is trivial, or, in the context of a more complex model, by using Markov chain Monte Carlo on the joint space of sensitivity parameters and parameters from the observed data model.
In the next section, we show that logistic-mEF models largely avoid this difficulty.

\subsection{Mixtures of exponential family models}
We now show that sensitivity analysis with logistic specifications is especially convenient under the weak assumption that the observed data model belongs to the class of exponential family mixtures.
In particular, we review results of \citet{franks2016non} showing that the mixture of exponential families assumption makes the normalizing constant in \eqref{eq:logistic missing} analytically tractable.



\subsubsection{Single exponential family models}
We first model the observed data as an exponential family distribution with natural parameter $\eta_t(X)$, possibly depending on $X$, 
\begin{equation}
\label{eq:exp-log-obs}
f^{\rm obs}_{t}(Y(t) \mid T=t, X) = h(Y(t))g(\eta_t(X))e^{s(Y(t)) ' \eta_t(X)}
\end{equation}
%
A key result from \citet{franks2016non} shows that when the selection function is assumed to have the logistic form in \eqref{eqn:logistic selection}, the missing data distribution belongs to the same exponential family as the observed data. 

\begin{proposition}
\label{prop:expo_miss}
Assume that the observed data is an exponential family with density $f^{\rm obs}_{t}(Y(t) \mid T=t, X) = h(Y(t))g(\eta_t)e^{s(Y(t)) ' \eta_t(X)}$ with sufficient statistic $s(Y(t))$ and natural parameter $\eta_t(X)$.  Further, assume the probability of selection is logistic in that statistic, $f(T = 1 \mid Y(t), X) = \text{logit}^{-1}\hbox{}\{\alpha_t(X) + \gamma_t' s_t(Y(t))\}$, such that $\eta^*_t(X) = \eta_t(X) + \gamma_t$ lies in the natural parameter space of the exponential family of $f^{\rm obs}_t$.
Then the distribution of the missing potential outcomes in arm $t$ is in the exponential family as $f^{\rm obs}_t$, and has density 
\begin{equation}
f^{\rm mis}_{\gamma_t, t}(Y(t) \mid T = 1\-t, X) = h(Y(t)) g(\eta^*_t(X)) e^{{s(Y(t))}'\eta^*_t(X))}.\label{eq:expo_miss}
\end{equation}
\end{proposition}

Proposition~\ref{prop:expo_miss} follows because, with an exponential family outcome model, the normalizing constant in Equation \eqref{eq:logistic missing} is analytically tractable, with $C(\gamma_t, X) = \frac{g(\eta_t(X))}{g(\gamma_t + \eta_t(X))}$.
In addition, the constraint that $\eta^*_t(X)$ lie in the natural parameter space ensures that the missing data distribution is proper.
Given the missing data distribution in \eqref{eq:expo_miss}, the implied complete data density $f_{\gamma_t, t}(Y(t))$ can then be expressed as a simple mixture of the observed and missing components.


%
%

To provide some intuition for how the factorization operates under the logistic selection with exponential families, we present two simple examples.

\begin{example}[Binary observed outcomes]
\label{ex:bernoulli}
Let $Y(t) \in \{0, 1\}$ and
\begin{align*}
f^{\rm obs}_t(Y(t) \mid T=t, X) &\sim \Bern(\text{Logit}^{-1}(\mu_t(X))\\
f(T \mid X, Y(t)) &\propto \Bern(\text{Logit}^{-1}(\alpha_t(X) + \gamma_t Y(t)))
\end{align*} 
\noindent By Proposition \ref{prop:expo_miss} the unobserved potential outcome distribution is:
\begin{align*}
f^{\rm mis}_{\gamma_t, t}(Y(t) \mid T=1\-t, X) &\sim \text{Bern}(\text{Logit}^{-1}(\mu_t(X) + \gamma_t))
\end{align*}
\noindent Thus, the extrapolation factorization with logistic treatment assignment applied to Bernoulli data implies an additive shift in the log-odds of the unobserved potential outcomes, relative to the observed potential outcome distribution.
\end{example}

\begin{example}[Normal observed outcomes] Let $Y(t)$ follow a Normal distribution. Because the Normal distribution is a two parameter exponential family distribution with sufficient statistics $s(Y(t)) = (Y(t), Y(t)^2)$, we consider a treatment assignment mechanism that is quadratic in the potential outcomes. 
\begin{align*}
f^{\rm obs}_t(Y(t) \mid T=t, X) &\sim \text{N}(\mu_t(X), \sigma_t^2)\\
f(T \mid Y(t)) &\propto \text{Bern}(\text{Logit}^{-1}(\alpha_t(X) + \gamma_t Y(t) + \psi_tY(t)^2))
\end{align*} 
\noindent By Proposition~\ref{prop:expo_miss} the unobserved potential outcome distribution is:
\begin{align}
\label{eqn:norm_example_mis}
f^{\rm mis}_{(\gamma_t, \psi_t), t}(Y(t) \mid T=1\-t, X) &\sim \text{N}\left(\frac{\mu_t(X) + \gamma_t\sigma_t^2}{1 - 2\psi_t\sigma_t^2} , \frac{\sigma_t^2}{1 - 2\psi_t\sigma_t^2}\right)
\end{align}
This model has two sensitivity parameters for each treatment arm. Assuming that the logistic function is linear in $Y(t)$, i.e. $\psi_t=0$, implies that standard deviations of observed and missing potential outcome distributions are identical.\footnote{Note that in the quadratic model we require $\psi_t > \frac{1}{2\sigma_t^2}$ to ensure propriety of the missing potential outcome distribution (Condition \ref{cond:int_contraint}).}
In Section \ref{sec:analysis_nhanes}, we estimate $\mu_t(X)$ from the observed data using BART, which allows us to conduct a sensitivity analysis on potential outcomes with flexibly estimated response surfaces. 
\end{example}

\subsubsection{Extension to mixtures}
We now extend the formulation to the full class of logistic-mEF models, which include outcome models that are mixtures of exponential family distributions.
Let 
$$f^{\rm obs}_t(Y(t) \mid \eta_t, T=t) = \sum_k \pi_k h_k(Y(t))g_k(\eta_{tk})e^{s(Y(t)) ' \eta_{tk}}$$
be a mixture of exponential family distributions with mixture weights $\pi_k$ and common sufficient statistics, $s(Y(t))$.  Then, the normalizing constant in Equation \eqref{eq:logistic missing} is analytically tractable with,  $C(\gamma_t, X) = \sum_k \pi_k \frac{g_k(\eta_{tk})}{g_k(\gamma_t+\eta_{tk})}$.  This leads to the following proposition:
\begin{proposition}
\label{prop:expo_miss_mix}
Assume that the observed data follows a $K$-component mixture exponential family distribution with density $f^{\rm obs}_t(Y(t) \mid \eta_t, T=t) = \sum_k \pi_k h_k(Y(t))g_k(\eta_{tk})e^{s(Y(t)) ' \eta_{tk}}$, with sufficient statistic, $s(Y(t))$, common across all components. Further, assume the probability of selection is logistic in the sufficient statistic, $f(T = 1 \mid Y(t), X) = \text{logit}^{-1}\hbox{}\{\alpha_t(X) + \gamma_t' s_t(Y(t))\}$, such that, for each component $k$, $\eta^*_{tk}(X) = \eta_{tk}(X) + \gamma_t$ lies in the natural parameter space of the exponential family.  Then the distribution of the missing potential outcomes in arm $t$ is also a $k$-component mixture of the same exponential family distributions and has density 
\begin{equation}
\label{eqn:ef_mix_miss}
f^{\rm mis}_{\gamma_t, t}(Y(t) \mid \eta_t, \gamma_t, T=1-t) = \sum_k \pi^*_k h_k(Y(t))g_k(\eta^*_{tk})e^{s(Y(t)) ' \eta^*_{tk}}
\end{equation}
where $\eta^*_{tk} = \eta_{tk} + \gamma_t$, and
\begin{align}
\label{eqn:ef_mix_weights}
\pi_k^* &= \frac{\pi_k\frac{g_k(\eta_k)}{g_k(\eta_k^*)}}{\sum_k^K
        \pi_k\frac{g_k(\eta_k)}{g_k(\eta_k^*)}}.
\end{align}
\end{proposition}

\noindent Consider the following simple example.

\begin{example}[Normal mixture observed outcomes]
\label{ex:mix_normal_dists}
We extend the Normal model above to a finite mixture of Normal distributions: 
$$f^{\rm obs}_t(Y(t) \mid T=t, X) \sim \sum \pi_k \text{N}(\mu_{tk}(X), \sigma_{tk}^2).$$
For simplicity we assume $\psi_t = 0$.  Then, by Proposition~\ref{prop:expo_miss} and Equation \eqref{eqn:ef_mix_weights}, 
\begin{align}
\label{eqn:mix_normal_dists}
f^{\rm mis}_{\gamma_t, t}(Y(t) \mid T=1-t, X) &\sim \sum \pi^*_k \text{N}(\mu_{tk}(X) + \gamma_t\sigma_{tk}^2, \sigma_{tk}^2)\\[0.5em]
\pi^*_k &\propto \pi_k \exp\frac{1}{2}\left({\frac{\mu_{tk}^2(X)}{\sigma_{tk}^2} - \left(\frac{\mu_{tk}(X)}{\sigma_{tk}} - \gamma_t\right)^2}\right). \nonumber
\end{align}
In this model, the sensitivity parameters $\gamma_t$ affect both the mixture weights and the component means: the mixture weight for component $k$ increases as $\gamma_t$ approaches $\frac{\mu_{tk}(X)}{\sigma_{tk}}$.  
\end{example}

In general, we can apply Tukey's factorization with logistic selection to observed data densities modeled with non-parametric Bayesian methods like Dirichlet process mixtures of any exponential family distribution, and can can easily adapt the mixture results to model structured semi-continuous data as well.  In Section \ref{sec:analysis_jtpa}, we demonstrate both of these features by modeling zero-inflated income data, and use a DPM model for the continuous component.

\section{Calibrating Sensitivity Parameters}
\label{sec:calibration}

Since  sensitivity parameters are not identified from the data, calibrating the magnitude of the parameters requires reasoning about plausible values using prior knowledge and domain expertise. 
We first discuss how to interpret the sign of the sensitivity parameters. We then turn to the magnitude of the sensitivity parameters and introduce a method to calibrate this quantity using information from observed covariates.

\subsection{Interpreting the sign of sensitivity parameters}
\label{sec:sign_interpretation} 

To interpret the sign of sensitivity parameters, we return to the logistic selection model and consider the case where $s(Y(t)) = Y(t)$; that is, the sufficient statistic in the logistic selection model is simply $Y(t)$ itself.
In this setting, $\gamma_0$ and $\gamma_1$ are scalars and the probability of assignment to treatment  is $f(T=1 \mid Y(t), X) = \text{logit}^{-1}( \alpha_t(X) + \gamma_t Y(t)))$.  


With this setup, the sensitivity parameters $(\gamma_0, \gamma_1)$ have a relatively straightforward interpretation: they specify how sufficient statistics of the potential outcomes are over- or under-represented among observed control and treated units. 
For example, $\gamma_1 > 0$ implies that units with large $Y(1)$ are over-represented among treated units, and thus the observed treated unit average will be larger than the average of the unobserved treatment outcomes. Likewise, $\gamma_0 > 0$ implies that units with large values of $Y(0)$ are more likely to be assigned to treatment, so the observed control unit average will be smaller than the average of the unobserved control outcomes.
%
We give several examples to illustrate interpretation in practice.

\begin{example}[Same Sign]
\label{ex:same sign}
Consider a study of a medical treatment with health outcome $Y$ and unmeasured income variable $U$.
Suppose that larger values of $Y$ correspond to good outcomes, that treatment is expensive, so $P(T = 1 \mid U)$ is increasing in $U$, and that higher income induces better outcomes, such that $Y(t) = \alpha + \tau t + \psi U + \epsilon$ with $\psi > 0$.
In this study, both $Y(0)$ and $Y(1)$ are positively correlated with $T$, so large values of $Y(1)$ are over-represented among the observed treated units, and large values of $Y(0)$ are under-represented among control units, corresponding to positive $\gamma_1$ and $\gamma_0$.  
When $\gamma_0$ and $\gamma_1$ have the same sign, the treatment and control group means are biased in opposite directions, so the ATE changes rapidly as the magnitude of the sensitivity parameters increases. 
\end{example}

\begin{example}[Opposite Signs]
Consider the canonical ``perfect doctor'' example \citep{rubin2003basic}, where a particular medical treatment is prescribed by a doctor who is able to perfectly predict $Y(0)$ and $Y(1)$ for each patient, and assigns the patient to whichever arm gives them the higher outcome.
Suppose that in this population $Y(1) - Y(0)$ is independent of $[Y(1) + Y(0)] / 2$; that is, the individual treatment effect for each unit is unrelated to each unit's expected outcome under random assignment to treatment.
For example, suppose that $[Y(0), Y(1)]$ are independently normal.
Then the control arm and the treated arm both over-represent high outcomes, so $\gamma_0 < 0$ and $\gamma_1 > 0$. 
When $\gamma_0$ and $\gamma_1$ have opposite signs, the treated and control group means are biased in the same direction, so the ATE changes slowly as the magnitude of the sensitivity parameters increases.
\end{example}

\begin{example}[Single-Arm Confounding]
\label{ex:single arm}
Consider a study evaluating a job training program with wage outcome $Y$. Prior to the study, a subset of units is randomly given access to an alternative program that they can attend if they do not enroll in the treatment. Let $U = 1$ indicate access to this alternative program and that units with access to the alternative program are less likely to enroll in treatment, so $P(T = 1 \mid U)$ is decreasing in $U$. 
Suppose that
the alternative program is beneficial on average, for example, suppose the potential outcomes are given by $Y(0) = \alpha(X) + \psi U + \epsilon$ with $\psi > 0$ and $Y(1) = \alpha(X) + \tau + \epsilon$.
Then, the observed outcomes under treatment are representative of the distribution of $Y(1)$, so $\gamma_1 = 0$, but, under control, units with higher wage outcomes are over-represented, so $\gamma_0 < 0$.
When only one of the sensitivity parameters is nonzero, only one of the group means is biased, so the ATE changes moderately when the magnitude of the nonzero sensitivity parameter increases. \end{example}

\subsection{Calibrating the magnitude of sensitivity parameters}
\label{sec:mag_calibration} 

\subsubsection{Calibrating against variation explained}
We now turn to the more challenging task of calibrating the magnitude of sensitivity parameters. 
Our primary proposal is to calibrate the magnitude of sensitivity parameters to the amount of variation in the treatment assignment $T$ that is explained by $Y(t)$, \textit{above and beyond what is accounted for by $X$}.%
\footnote{One evident approach, which we do not endorse here, is to reason about the magnitude of $\gamma_t$ directly, calibrating it against regression coefficients inferred from observed covariates \citep{dorie2016flexible, blackwell2014selection, middleton2016bias}.  The problem with this approach is that it can be difficult to interpret coefficients due to collinearity between and among $X$ and $Y(t)$; see, for example, \citet{oster2017unobservable}. In contrast, our approach is robust to multicollinearity among the predictors. See \citet{cinelli2018making} for additional discussion.}
To instantiate this idea, we adopt a version of the ``implicit $R^2$'' measure from \citet{imbens2003sensitivity}, which generalizes variance-explained measures to the case of logistic regression.  This approach is consistent with corresponding proposals for calibration in some latent confounder models \citep{imbens2003sensitivity, cinelli2018making}, with the important difference that our parameterization involves calibrating the variance explained by the potential outcomes rather than by a latent confounder.  

To fix ideas, we will again focus on the simple case where the sufficient statistic is $s(Y(t)) = Y(t)$ and $\gamma_0$ and $\gamma_1$ are scalars, so that following Equation \eqref{eqn:logistic selection}, the probability of assignment to treatment  is $f(T=1 \mid Y(t), X) = \text{logit}^{-1}( \alpha_t(X) + \gamma_t Y(t)))$.  
In the discussion that follows, we work on the logit scale and let $m(X) = \mathrm{logit}(e(X))$ be the logit of the propensity score given variables $X$. To characterize variance explained, we first note that the logistic treatment model can be expressed using a latent variable formulation as
\begin{align*}
Z &= m(X) + \epsilon \text{ with } \epsilon \sim \text{logistic}(0, 1)\\
T &= 
\begin{cases}0 \text{ if } Z < 0\\
1 \text{ if } Z \geq 0
\end{cases}
\end{align*}


\noindent Since $T$ is a deterministic function of the latent $Z$, it is sufficient to characterize how well $Y(t)$ and $X$ explain $Z$.  Because the variance of a standard logistic is $\pi^2/3$, the total variance of $Z$ is simply $\text{Var}(m(X)) + \pi^2/3$.  We then define the variance explained by $X$, and the partial variance explained by $Y(t)$ given $X$, respectively, as
\begin{align}
\rho^2_X &= \frac{\text{Var}(m(X))}{\text{Var}(m(X)) + \pi^2/3}, &0 \leq \rho^2_X \leq 1&\\
\rho^2_{Y(t) \mid X} &= \frac{\rho^2_{X, Y(t)} - \rho^2_{X}}{1-\rho^2_{X}}, &0 \leq \rho^2_{Y(t) \mid X} \leq 1.&
\end{align}
The partial variance explained by $Y(t)$ given $X$, $\rho^2_{Y(t) \mid X}$, represents the fraction of previously unexplained variance in $T$ that can now be explained by adding $Y(t)$ to the set of predictors $X$.  

With our method, we propose a target value, $\rho^2_*$, for the the unidentified $\rho^2_{Y(t) \mid X}$. Although the decision ultimately falls to domain expertise, we suggest using observed predictors to set the target $\rho^2_*$ by analogy.
For example, for each covariate $X_j$, we can compute a partial variance explained by $X_j$ given all other covariates $X_{-j}$, and  set the target $\rho^2_* = \rho^2_{X_j \mid X_{-j}}$ for an appropriate covariate based on expert knowledge.  We interpret this to mean that the information gained by adding $Y(t)$ to $X$ as a predictor of treatment assignment is comparable to the information gained by adding $X_j$ to $X_{-j}$.  We give concrete examples of this approach in the analysis in Section \ref{sec:analysis_nhanes}.
In some cases, alternative calibration schemes may be more appropriate.  For example, one could calibrate $\rho^2_{Y(t) \mid X_A}$ for some core subset of covariates $X_A \subset X$ using \eqref{eqn:calibrator}, replacing $X$ with $X_A$.
In Section~\ref{sec:analysis_jtpa}, we take this approach and apply a calibration scheme setting $X_A$ to be the empty set.
See \citet{cinelli2018making} for a discussion of related alternatives in the context of linear modeling.

\subsubsection{Mapping between variation explained and sensitivity parameters}
Once a value of $\rho^2_*$ has been chosen, we must then identify the magnitude of the corresponding sensitivity parameter, $\gamma_t$.  Since $\rho^2_{Y(t) \mid X}$ is monotone in $\gamma_t$, there is a one-to-one mapping between the two quantities that can be used to identify $\gamma_t$.  We formalize this in the following proposition: 

\begin{proposition}[Calibration Identity]
\label{prop:calibration}
Suppose that
$\text{logit}(e(X, Y(t) ))$ is linear in $Y(t)$, with the form in \eqref{eqn:logistic selection}. Further, suppose that $\text{logit}(e(X)) = m(X)$ and $\sigma_{rt} = \sqrt{E[Var(Y(t)\mid X)]}$.\footnote{In homoscedastic models, $\sigma_{rt}$ is simply the residual standard deviation of the potential outcome $sd(Y(t) \mid X)$, and is independent of $X$.}
Then,
\begin{align}
\label{eqn:partial_r2}
\rho^2_{Y(t) \mid X} = 
\frac{\sigma^2_{rt} \gamma_t^2}{{\rm Var}(m(X)) + \pi^2/3 +  \sigma^2_{rt} \gamma_t^2},
\end{align}
\noindent The inverse of this equation identifies the magnitude of $\gamma_t$:
\begin{equation}
\label{eqn:calibrator}
|\gamma_t| = \frac{1}{\sigma_{rt}}\sqrt{\frac{\rho^2_{Y(t) \mid X}}{1 - \rho^2_{Y(t) \mid X}}({\rm Var}(m(X)) + \pi^2/3}).
\end{equation}
\begin{proof}
See appendix.
\end{proof}
\end{proposition}

\noindent We use \eqref{eqn:calibrator}  to translate any fixed target on $\rho^2_*$ to a value for $|\gamma_t|$.  Note that \eqref{eqn:calibrator} only requires estimates of ${\rm Var}(m(X))$ and $\sigma_{rt}$.  Estimating ${\rm Var}(m(X))$ is trivial using any propensity score model, but estimating the residual standard deviation $\sigma_{rt}$ is slightly more nuanced. This is because $\sigma_{rt}$ is a property of the complete distribution of the potential outcome, $Y(t)$, which itself depends on $\gamma_t$.  We write $\sigma_{rt}(\gamma_t)$ to emphasize that the residual standard deviation depends on $\gamma_t$; $\sigma_{rt}(\gamma_t)$ is available in analytic form for the mixture of exponential family models considered in Section~\ref{sec:logit-ef}.  We then numerically solve 
$$\sigma_{rt}(\gamma_t)|\gamma_t|= \sqrt{\frac{\rho^2_{Y(t) \mid X}}{1 - \rho^2_{Y(t) \mid X}}({\rm Var}(m(X)) + \pi^2/3})$$
\noindent for $\gamma_t$.  In many of observed-data model specifications discussed in this paper, the residual standard deviation in both the observed and missing data densities are identical, so this recursive identification of $\gamma_t$ is not necessary (e.g. as in Equation \ref{eqn:norm_example_mis} for $\psi_t = 0$).  

%% file: sections/sec-analysis.tex
\section{Applications of the Logistic-mEF Model}
\label{sec:analysis}

We now apply our approach to two examples. 
In the first example, we conduct sensitivity analysis for 
the effects of blood pressure medication.  In this example, we use a nonparametric estimate of the response surface but assume normally distributed residuals.  
In the second example, we conduct sensitivity analysis on the effect of a job training program on income and employment. In this example, the outcome is zero-inflated, and we focus on quantile treatment effects rather than average treatment effects.  In addition to demonstrating the flexibility in modeling structured data, we also show how Tukey's factorization can be applied to nonparametric residual models, by modeling the continuous component of the observed data densities using a Dirichlet Process Mixture of normal distributions.  
In both examples, we demonstrate the flexibility of our approach by estimating the posterior distribution of quantile treatment effects for a range of quantiles.

\subsection{Analysis of NHANES data}
\label{sec:analysis_nhanes}
We consider a study aimed at estimating the effect of `taking two or more anti-hypertensives' on average diastolic blood pressure using data from the Third National Health and Nutrition Examination Survey (NHANES III) \citep{NHANES3}, a comprehensive survey of Americans’ health and nutritional status.  We follow the same set up as  \citet{dorie2016flexible}, and utilize pre-treatment covariates like race, gender, age, income, body mass index (BMI), and whether the patient was insured, among others.  We let $Y(t)$ be the average diastolic blood pressure for a subject in treatment arm $t$, where $t=1$ means the subject was taking two or more anti-hypertensive medications and $t=0$ means the subject was not.    

In this application, we show that our method works well with a nonparametric model for the response surfaces.
%
First, we assume the following data generating model for the potential outcomes: 
\begin{align}
f^{\rm obs}_t(Y(t) \mid T=t, X) &\sim N(\mu_t(X), \sigma_t^2)\\
(\mu_t(X), \sigma_t^2) &\sim \text{BART}(X \mid T=t) \label{eqn:bart_indep}\\
f(T \mid Y(t), X) &\sim \text{Bern}(\text{Logit}^{-1}(\alpha_t(X) + \gamma_tY(t)))
\end{align} 
%
\noindent As shown in Section \ref{sec:logit-ef}, the missing data distribution is therefore:
\begin{equation}
f^{\rm mis}_{\gamma_t, t}(Y(t) \mid T=1\-t, X) \sim N(\mu_t(X) + \gamma_t\sigma_t^2, \sigma_t^2).\\
\end{equation}
Again following \citet{dorie2016flexible} we use Bayesian Additive Regression Trees \citep[BART; ][]{chipman2010bart} to flexibly model $\mu_t(X)$ and $\sigma_t^2$. In contrast to their approach, we focus on using BART to estimate only the \emph{observed} potential outcomes and do not incorporate latent confounders into the BART model.  
We use independent BART prior distributions for the mean surface, $\mu_t(X)$, for each treatment arm and similarly model separate residual variances for each arm \eqref{eqn:bart_indep}.
The ATE and QTE can then be estimated as functions of the estimated marginal complete data distributions for the potential outcomes (averaged over covariates) where:
\begin{align}
\label{eq:nhanes-complete}
f(Y(t)) &\sim \sum_{i=1}^{N_t} \frac{1}{N_t} N(\mu_t(X_i), \sigma_t^2)+ \sum_{j=1}^{N_{1\-t}} \frac{1}{N_{1\-t}} N(\mu_t(X_j) + \gamma_t\sigma_t^2, \sigma_t^2).
\end{align}

\paragraph{Calibration.} 
We calibrate the magnitude of the sensitivity parameters using the approach outlined in Section \ref{sec:calibration}.  
We illustrate this approach with body mass index (BMI), one of the most important predictors in terms of partial variance explained, with $\rho^2_{X_{\rm{BMI}} \mid X_{-\rm{BMI}}} \approx 0.04$.
To map this value to sensitivity parameters, we use the estimated residual standard deviation of the potential outcomes in each arm,  $\hat{\sigma}_{r0} \approx 11.2$ and $\hat{\sigma}_{r1} \approx 10.9$. We can then apply the formula in Equation \eqref{eqn:calibrator} to obtain $|\gamma_0| \approx 0.046$ and $|\gamma_1| \approx 0.048$.\footnote{Specifically, the formula in \eqref{eqn:calibrator} implies $\sigma_{rt} \gamma_t \approx 0.52$. In turn, $|\gamma_0| \approx 0.52 / 11.2 \approx 0.046$ and $|\gamma_1| \approx 0.52 / 10.9 \approx 0.048$.}
In general, we limit our sensitivity analysis to unmeasured confounders up to this magnitude.

\begin{figure}	
	\centering
	\begin{subfigure}[t]{0.49\textwidth}
		\includegraphics[width=\textwidth]{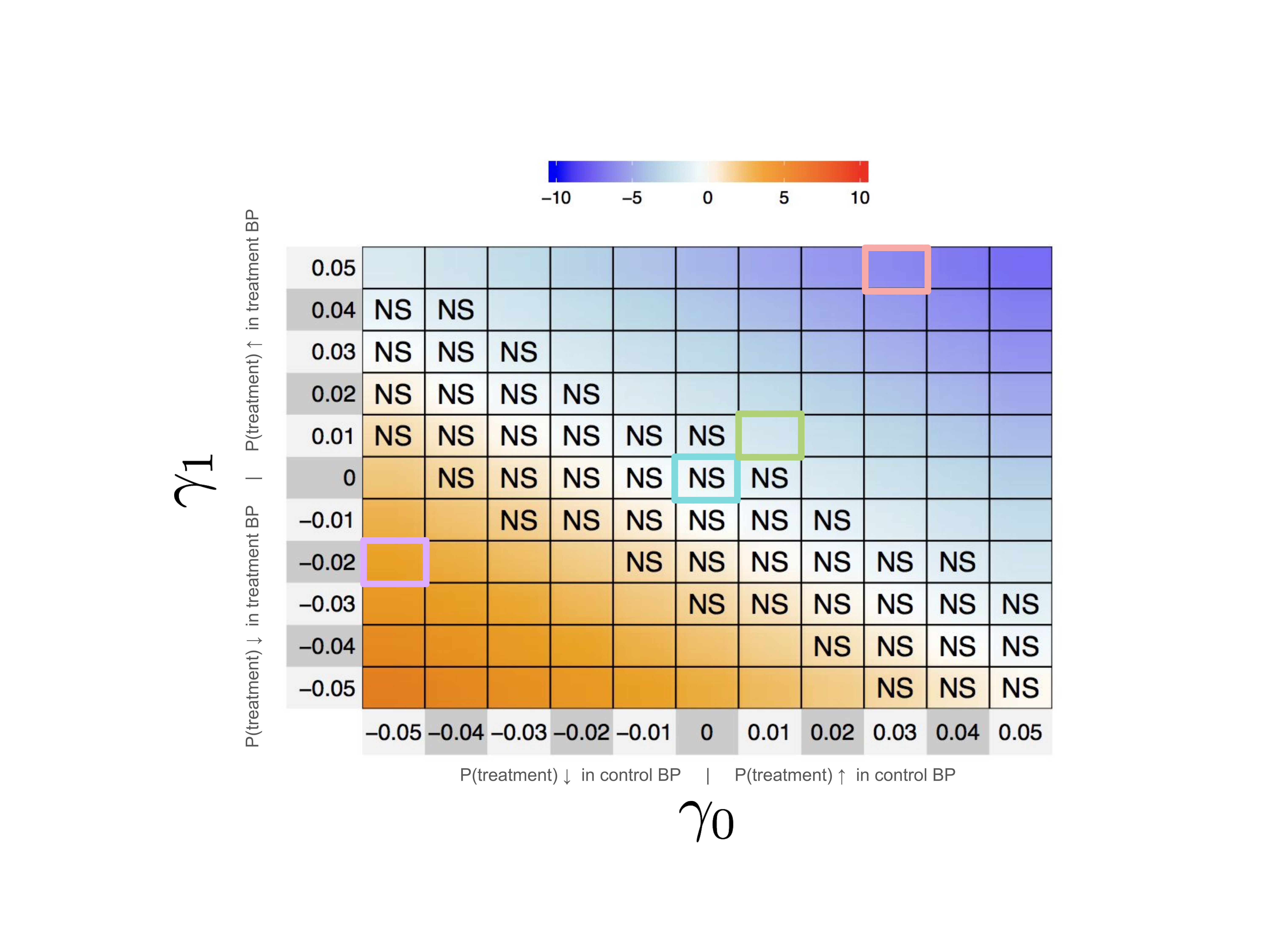}
		\caption{ATE}\label{nhanes_interaction}
	\end{subfigure}
    \hspace{.3in}
		\begin{subfigure}[t]{0.4\textwidth}
		\centering
		\includegraphics[width=\textwidth]{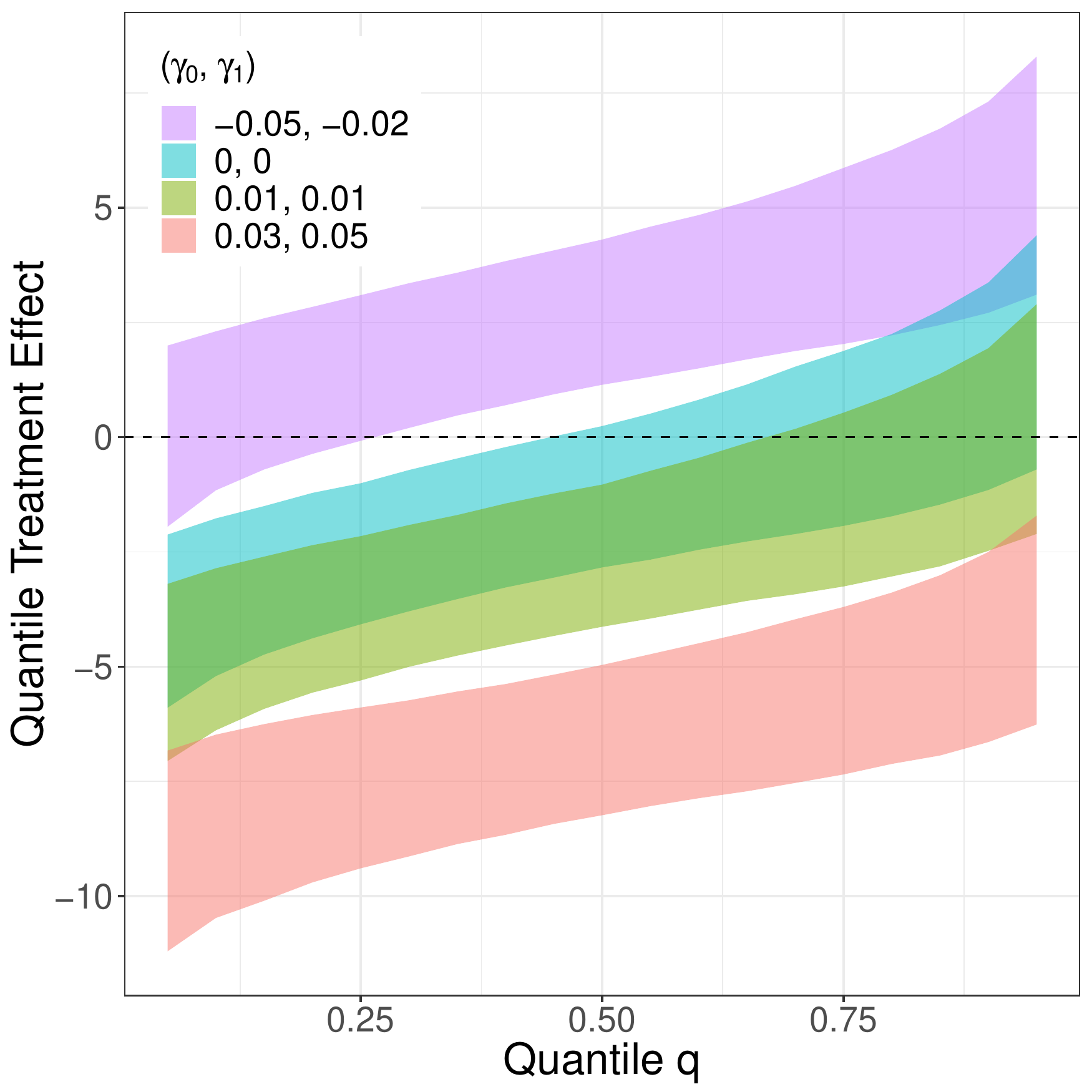}
		\caption{Quantile effects}\label{fig:nhanes_qte}		
	\end{subfigure}	

	\caption{Average and quantile treatment effects of diuretics on diastolic blood pressure. Treatment effect measured in units of millimeters of mercury (mmHg). NS denotes ``not significant''. a) Estimated ATEs for the BART model. Under unconfoundedness, the ATE is negative but not significant.  c) Quantile treatment effects, $\tau_q$. $\tau_q$ is increasing in $q$ because the treatment potential outcomes have higher variance than the control potential outcomes.  Colored boxes in (a) correspond to QTE distributions in (b).}
	\label{fig:1}
\end{figure}

\paragraph{Results.} 
In Figure \ref{nhanes_interaction} we visualize ATE estimates for a grid of sensitivity parameters.  Here ``NS'' denotes ``not significant'', by which we mean the 95\% posterior credible interval of the ATE contains 0.  Although the notion of ``significance'' vs ``non-signifiance'' is fragile, it still provides a measure of the uncertainty associated with the estimated effects; see \citet{dorie2016flexible}. 

Under unconfoundedness ($\gamma_0 = \gamma_1 = 0$) the posterior mean for the ATE is approximately -0.7 mmHg but there is enough posterior uncertainty that the effect is not significantly different from 0 (light blue box in Figure \ref{nhanes_interaction}). The ATE changes the most along the diagonal parallel to $\gamma = \gamma_0 = \gamma_1$, with the ATE no longer significantly different from 0 when $\gamma = 0.01$. Figure~\ref{fig:1} also highlights the sensitivity patterns discussed in Examples~\ref{ex:same sign}--\ref{ex:single arm}, where the ATE is far more sensitive to the magnitudes of $\gamma_0$ and $\gamma_1$ when they have the same sign.  For example, for the pink box, with $\gamma_0 = 0.03$ and $\gamma_1 = 0.05$, the posterior mean ATE is approximately $-5.6$ mmHg, which is roughly half a (marginal) standard deviation larger than the estimate under unconfoundedness.
For these sensitivity parameter settings, the variance explained by the control potential outcome is comparable to the variance explained by the indicator for having health insurance, that is $\rho^2_{Y(0) \mid X} \approx \rho^2_{X_{\rm{ins}} \mid X_{-\rm{ins}}}$; the variance explained by the treatment potential outcome is comparable to the variance explained by BMI, that is $\rho^2_{Y(1) \mid X} \approx \rho^2_{X_{\rm{BMI}} \mid X_{-\rm{BMI}}}$  (see Figure \ref{fig:nhanes_calibration_comparison}).   

In Figure \ref{fig:nhanes_qte}, we plot the 95\% posterior credible band for the quantile treatment effects, $\tau_q$, for several settings of the sensitivity parameters (colors match squares those in Figure \ref{nhanes_interaction}).  Interestingly, $\tau_q$ is increasing with $q$ for all combinations of $\gamma$, which occurs because the estimated residual variance for the treated potential outcomes is slightly larger than for the control potential outcomes.  Similarly, for the complete data, the difference between the largest and smallest treatment potential outcomes is larger than the difference in the control potential outcomes. This is consistent with a situation in which the treatment varies across individuals beyond what is explained by covariates \citep{ding2018decomposing}.

In the appendix, we also demonstrate the separation between the observed data model and the sensitivity analysis in our framework. Specifically, we show how results change when we use different observed data models with the same treatment selection specifications as above. We provide results for two additional observed data models: an alternative BART parameterization and the Bayesian Causal Forest (BCF) model \citep{hahn2017bayesian}.
Broadly, our results are consistent with \citet{dorie2016flexible}, who show that both effect size and significance in this example can be sensitive to changes in the outcome model (testable) and treatment selection (untestable).
Finally, while we use a flexible model to estimate the response surfaces, the quantile effects are still sensitive to the assumption of normality on the residuals. We turn to this next.

\subsection{Analysis of Job Training Data}
\label{sec:analysis_jtpa}

In this example, we conduct a sensitivity analysis for quantile treatment effects for zero-inflated income data.  Zero-inflated outcomes are common in a range of settings; we focus on the context of evaluating job training programs.  In these studies, the primary outcome of interest, income, is zero for individuals who are unemployed and thus the average treatment effect misses important variation \citep{heckman1997making, bitler2006mean}. As a result, several studies have instead focused on estimating quantile treatment effects in these settings.
Specifically, we consider the well-known study from \citet{abadie2002instrumental}, who estimate quantile treatment effects in the context of the Job Training and Partnership Act (JTPA) evaluation, a large randomized trial estimating the impact of select workforce development programs on wages.
In this analysis we focus only on individuals randomly assigned to treatment, and compare outcomes between those who choose to participate in the program ($T=1$) versus those who did not ($T=0)$.  We choose this artificial setup --- comparing participants and non-participants among those randomly assigned to treatment --- specifically because selection bias is a clear concern. 

\begin{figure}[tbp]
	\centering
	\begin{subfigure}[t]{0.45\textwidth}
		\centering
		\includegraphics[width=\textwidth]{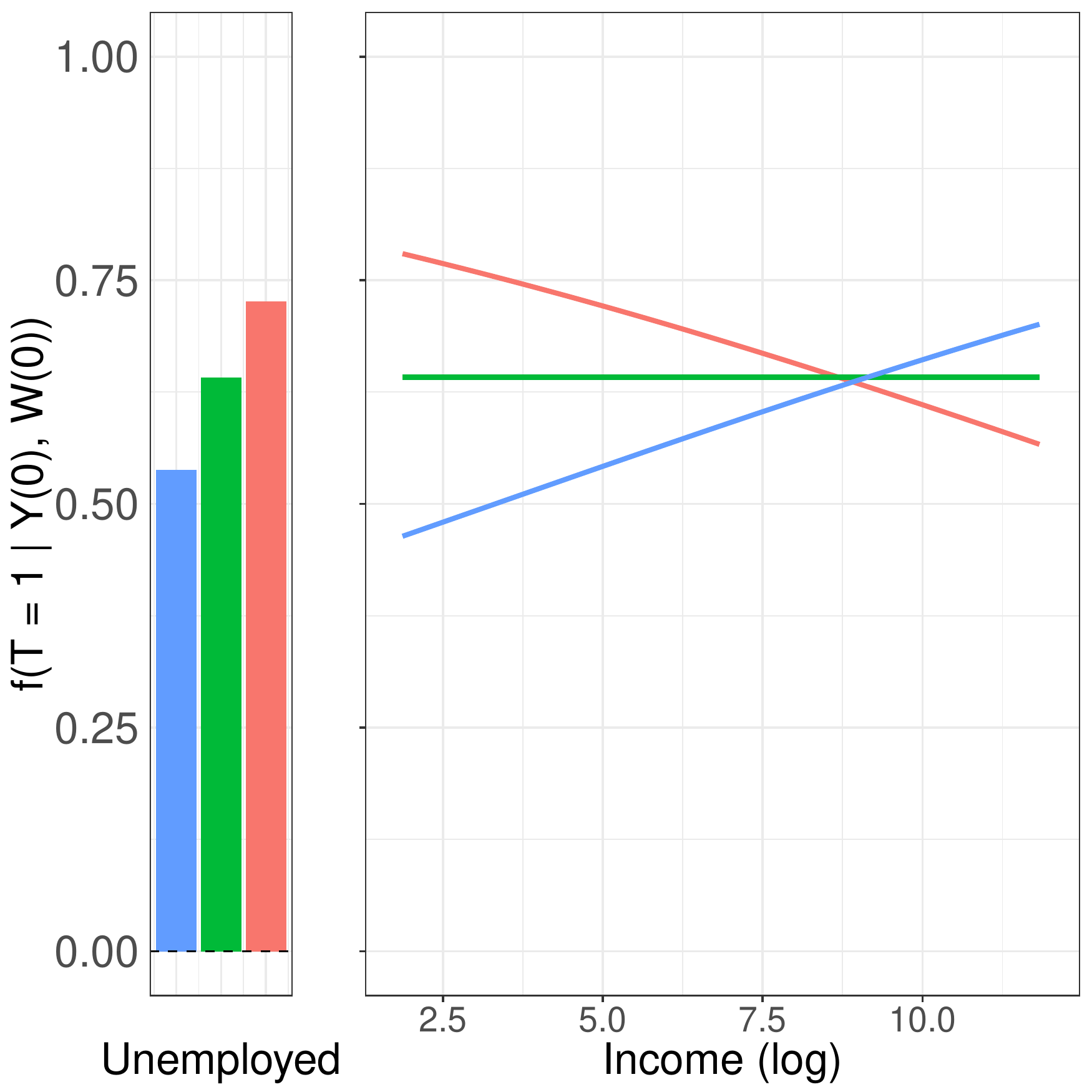}
		\caption{User-specified selection functions}\label{fig:job_training_treat_prob}		
	\end{subfigure}
	\\[2em]
	\begin{subfigure}[t]{0.45\textwidth}
		\centering
		\includegraphics[width=\textwidth]{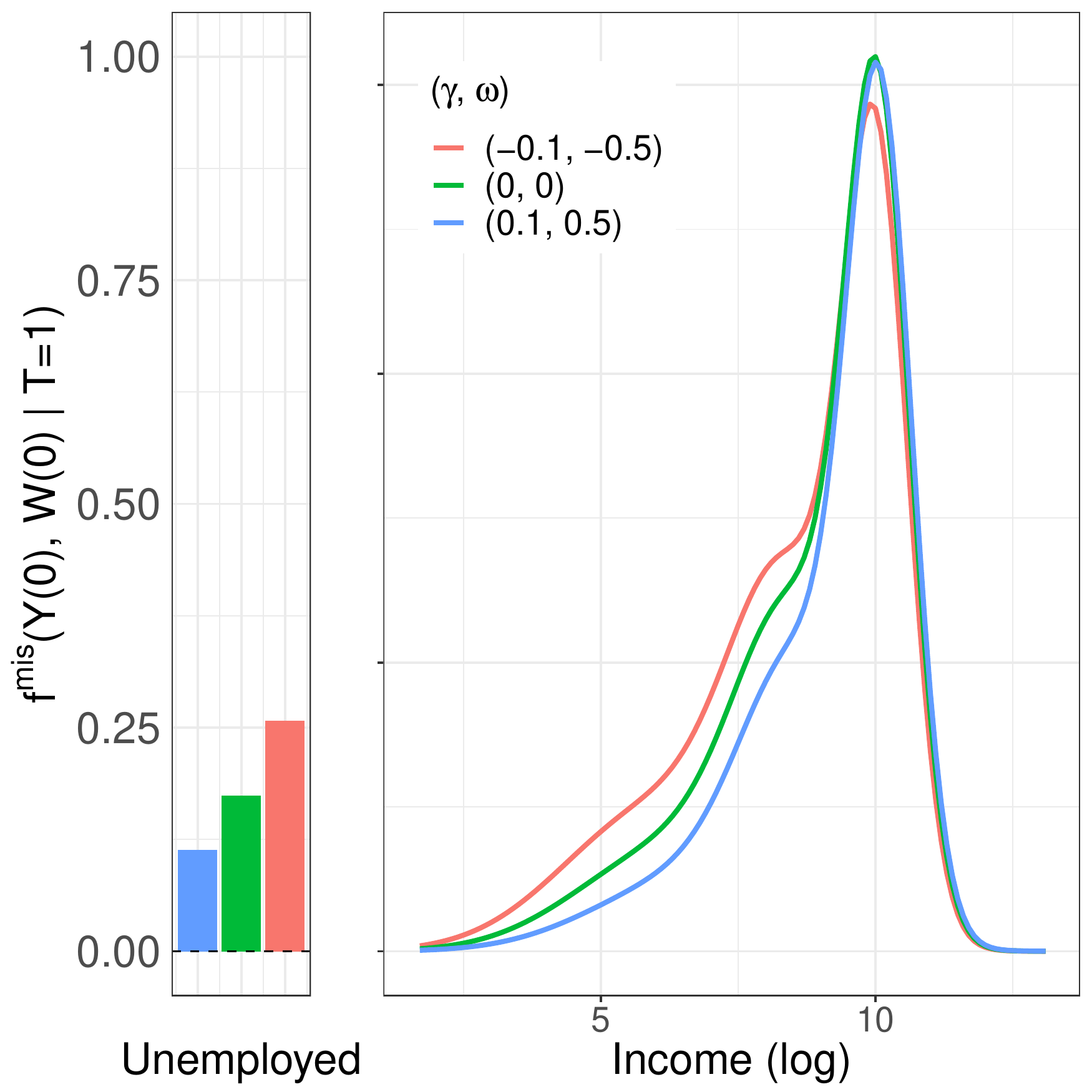}
		\caption{Missing control outcomes}\label{fig:job_training_missing}
	\end{subfigure}\qquad
		\begin{subfigure}[t]{0.45\textwidth}
		\centering
		\includegraphics[width=\textwidth]{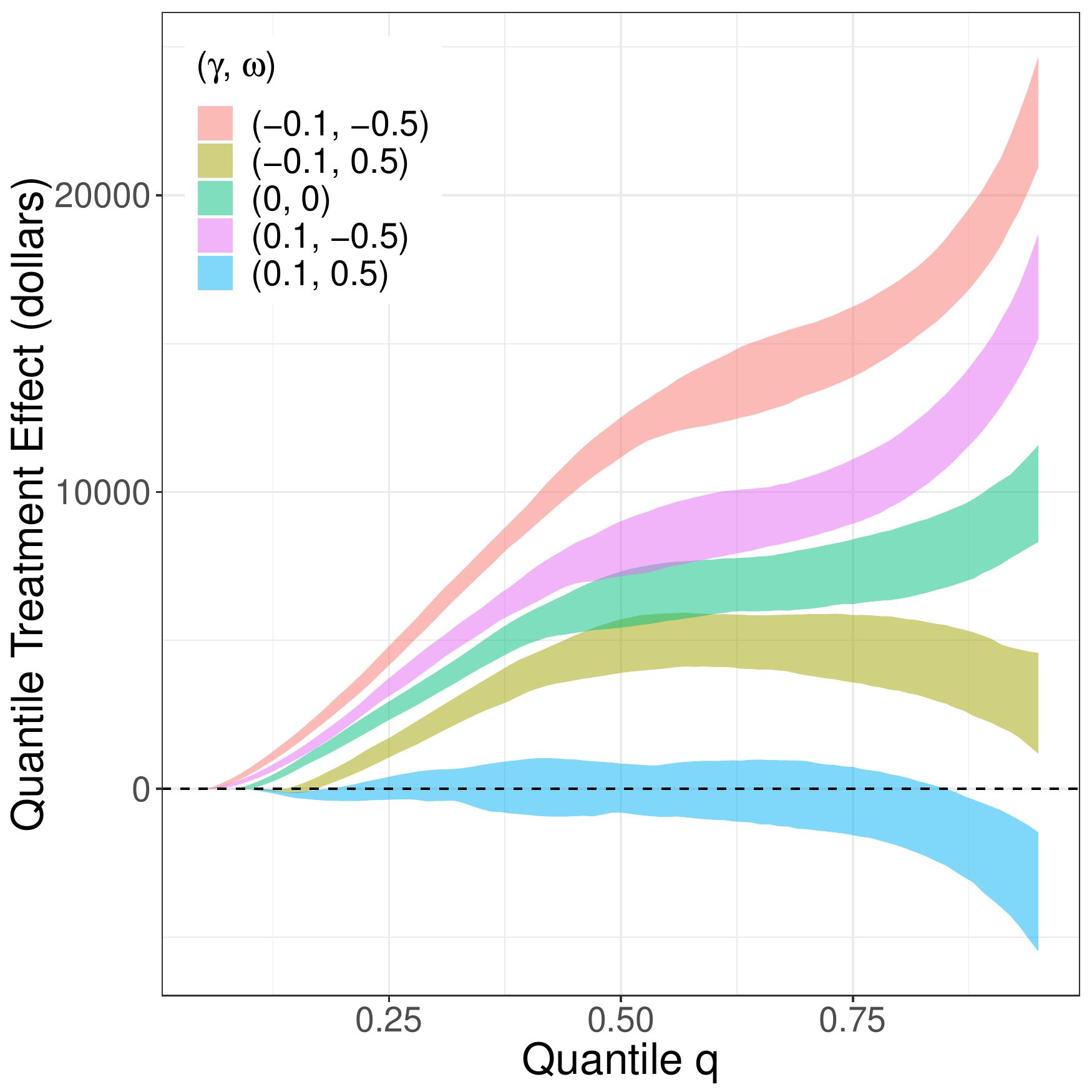}
		\caption{Quantile effects}\label{fig:job_qte}		
	\end{subfigure}

	\caption{Example model specifications for non-participators (a,b) and quantile treatment effects (c) for the JTPA data set.  a) Participation probability as a function of non-participator potential outcomes, $(Y(0), W(0))$.  The left bar plot shows participation probability given employment status, $f(T=1 \mid W(0))$ and the right depicts participation vs log income, $f(T=1 \mid \text{log}(Y(0)))$, for three different pairs of sensitivity parameters. b) The probability of unemployment, $E[W(0)\mid T=1]$, for the missing control units (left) and log income density for the employed among the missing control outcomes, $f(\text{log}(Y(0)) \mid T=1)$, (right). c) 95\% posterior confidence bands for $\tau_q$, for five pairs of sensitivity parameters.   \label{fig:jtpa}}
\end{figure}

%
%
%
This analysis is designed to showcase that our sensitivity framework allows investigators to conduct sensitivity analysis even when they employ flexible models of complex data.
Following previous work, we develop a two-part model for the semi-continuous data \citep{duan1983comparison, olsen2001two, javaras2003multiple}.  Let $Y(t)$ be income and $W(t) \in \{0, 1\}$ an indicator for employment status with $W(t) = 0 \implies Y(t) = 0$ and $W(t) = 1 \implies Y(t) > 0$. 
For simplicity, we exclude covariates in this analysis and focus on nonparametric estimation of the observed potential outcome distributions.   Specifically, we flexibly model the observed (log) income among the employed using a Dirichlet Process mixture of normal distributions \citep{neal2000markov}:
\begin{align}
f^{\rm obs}_t(\log(Y_i(t)) \mid T=t, W=1)   &\sim N(\mu_{it}, \sigma_{it}^2) \label{eqn:dp_spec1}\\
(\mu_{it}, \sigma_{it}) &\sim G\\
G &\sim DP(\alpha G_0)
\label{eqn:dp_spec3}
\end{align}
\noindent where $G$ is the conjugate normal-inverse gamma prior distribution for the normal likelihood.  To estimate the observed data density, we use the ``dirichletprocess'' R package, which implements a truncated stick breaking process to approximate the infinite mixture weights \citep{dirichletprocess}.

We then propose the following treatment selection specification, with separate selection functions for employed and unemployed individuals:
%
\begin{align}
f(T=t \mid \log(Y(t)), W(t)=1)   &\sim \text{logit}^{-1}(\beta_t + \gamma_t\log(Y(t))) \label{eqn:jtpa_sel1}\\
f(T=t \mid W(t) = 0)   &\sim \text{logit}^{-1}(\alpha_t + \omega_t I\{W(t)=0\}) 
\label{eqn:jtpa_sel2} 
\end{align}
In words, for employed individuals, the participation probability is logistic in the log income.  
Analogous to the Normal mixture example in Section \ref{sec:logit-ef} (Example \ref{ex:mix_normal_dists}), under the logistic selection model, the missing potential log-incomes $Y(t)$ also follow a Dirichlet Process mixture of Normals but with different component means and mixture weights.  
For unemployed individuals, the odds of participation increase multiplicatively by $\exp(\omega_t)$. Analogous to the Bernoulli example (Example \ref{ex:bernoulli}), under the logistic selection model, the missing potential employment outcome $W(t)$ is also Bernoulli but with an additive shift in the log-odds.




\paragraph{Calibration.} Reflecting the zero-inflated data structure, we handle calibration separately for employed and unemployed individuals. By way of demonstration, we use covariates to calibrate our sensitivity analysis but exclude them from treatment effect estimation.\footnote{This is not a viable approach in a real analysis without covariates, since there is no $X$ from which we can estimate $\rho^2_X$.}  The covariates, $X$, that we use for calibration are race, marital status, gender, age, and high school diploma or equivalent.

First, we calibrate the magnitude of $\gamma_t$, the sensitivity parameters for (log) income.  In this analysis, we focus on subset of the sensitivity parameter space in which $\gamma_0 = \gamma_1 = \gamma$.  We then calibrate $\gamma$ by fixing target values of $\rho^2_{Y(t)}$ to be approximately $\rho^2_X$, the partial variance explained by the vector of observed covariates.\footnote{$\rho^2_{Y(t)} = \rho^2_{Y(t) \mid \emptyset}$ is the marginal variance explained.}  We find that for the subset of employed individuals, the variance in $T$ explained by $X$ is $\rho^2_X \approx 0.01$.  By Equation \eqref{eqn:calibrator}, we find that $|\gamma_t| \approx 0.1$ when $\rho^2_{Y(t)} = 0.01$, where we use the fact that $\text{Var}(m(X)) = \text{Var}(m(\emptyset)) = 0$.

Second, we calibrate the magnitude of $\omega_t$, the sensitivity parameter for (binary) employment, $W(t)$.  We follow the same calibration strategy by setting the target for $\rho^2_{W(t)} \approx \rho^2_X$.  We find that for the subset of unemployed individuals, the variance in $T$ explained by $X$ is $\rho^2_{W(t)} \approx 0.015$.  Applying Equation \eqref{eqn:calibrator}, this value of $\rho^2_{W(t)}$ corresponds to a value of $|\omega_t| \approx 0.5$. 

Figure \ref{fig:job_training_treat_prob} visualizes different user-specified choices for the unidentified selection model (\ref{eqn:jtpa_sel1}--\ref{eqn:jtpa_sel2}), or probability of assignment to treatment, implied under three different settings of $(\gamma, \omega)$. Here, $\omega$ determines the height of the bar in the left panel, and $\gamma$ determines the steepness of the curve in the right panel. 
\paragraph{Results.}  
In this section, we summarize our results, as implied by the choice for this treatment assignment function and the observed data density estimated with the Dirichlet Process mixture model (\ref{eqn:dp_spec1}--\ref{eqn:dp_spec3}) in Figures \ref{fig:job_training_missing} and \ref{fig:job_qte}. First, in Figure \ref{fig:job_training_missing} we display the distribution of missing control outcomes for those units assigned to treatment, $f^{\rm mis}_{(\gamma, \omega),0}(Y(0), W(0) \mid T = 1)$.   
From this figure, it is clear that the observed data imply very different counterfactual employment and income distributions under different sensitivity parameter settings. The thickness of the lower tail of the income distribution for the employed appears to be particular sensitive to selection effects. The fraction of unemployed among the missing control outcomes assigned to treatment is roughly twice as large for $\omega = -0.1$ as it is for $\omega = 0.1$.  

Figure \ref{fig:job_qte} summarizes sensitivity in the QTEs, which are the primary objects of interest.
Consistent with the distributional differences in Figure \ref{fig:job_training_missing}, however, the QTEs are sensitive in terms of both employment and income levels.
First, under all sensitivity settings we explore, the QTE at the lowest quantiles are identically zero, since both control and treated outcome distributions have a point mass at zero.
The QTEs start to diverge from zero at different values
 $q$, which are related to the proportion unemployed in the treatment and control populations.
For example, for $\omega = 0$ or $-0.4$, the proportion of units who are unemployed under treatment is smaller than the proportion of units who are unemployed under control, $f(W(1)=0) < f(W(0)=0)$.
Here, the QTE increases away from 0 at $q = f(W(1) = 1)$.
By contrast, for $\omega = 0.4$, the proportion unemployed under treatment is actually larger than the proportion unemployed under control, $f(W(1)=0) > f(W(0)=0)$, and thus the QTE decreases away from 0 (briefly) at $q = f(W(0) = 1)$.
Differences in income effects are more straightforward to read from the plot.
For the sensitivity parameter settings used in this analysis, the income under treatment $Y(1)$ generally has larger income quantiles than the distribution of income under control $Y(0)$.  Only when selection effects are very large, e.g. $(\gamma, \omega) = (0.1, 0.4)$, do the effects lose significance and shift toward slightly negative values.  
In the end, the shape of the estimated QTEs is similar to those in \citet{abadie2002instrumental}, though the impacts are larger due to the artificial setup.

%% file: sections/sec-discussion.tex
\section{Discussion}
\label{sec:discussion}

In this paper, we proposed a framework for sensitivity analysis in causal inference employing Tukey's factorization.
The framework has a number of advantages. First, it cleanly separates the identified and unidentified portions of the data-generating process. This guarantees that sensitivity parameters are unidentified, in contrast to many latent confounder models, and decouples model checking and sensitivity analysis.
Second, it only requires the data to be fit once, reducing computational burden and enabling \textit{post facto} sensitivity analysis for a wide range of models that previously assumed  unconfoundedness.
Third, it supports intuitive sensitivity parameterizations that investigators can calibrate to selection on observed covariates.

Extensions of our framework could fit particularly well with modern causal inference workflows that employ a similar separation of observed data modeling and causal reasoning.
In these workflows, the analyst first focuses on optimizing a fit to the observed data distribution, often employing heuristics such as cross validation to select or combine models.
The analyst then plugs predictions from this model into an estimation step tailored to the estimand of interest \citep{van2011targeted, chernozhukov2016double, xu2018bayesian}.
A sensitivity analysis based on Tukey's factorization would allow investigators to assess sensitivity in this workflow without putting constraints on the flexible model used in the first stage.
In particular, following \eqref{eq:fmis extrapolation}, such a sensitivity analysis could be implemented by adding a weighting step, parameterized by sensitivity parameters, between the first and second stages.

Although we highlighted many different use cases, there are several extensions we did not explore.
First, we focused here on sensitivity specifications that are independent of the covariates, which enabled us to limit the number of sensitivity parameters.
A natural extension would generalize these specifications to infer causal effects under covariate-varying values of the sensitivity parameter, i.e. $\gamma_t(X)$ \citep[see][]{jung2018algorithmic}.
Such an extension would increase the number of sensitivity parameters, introducing several challenges in calibration and reporting results.
A second extension would generalize our approach to multiple or multi-level treatments \citep{imai2004causal}.
This would require generalizing the factorization in Equation \eqref{eqn:extrapolation_causal}.  
In some special cases, for example, where the unobserved confounding can be represented as latent factors of the observed distribution of treatments \citep{wang2018blessings}, more parsimonious sensitivity factorizations may be possible, e.g., following \citet{damour2019aistats}.
A final extension would extend our sensitivity analysis to observational studies with missing data.
Specifically, we could combine approaches for dealing with informative dropout, previously applied in experimental settings \citep{Daniels2000, scharfstein2003incorporating}, with the models for confoundedness in observational studies described in this paper. 
We leave it to future work to explore these and other classes of useful extrapolation models, including models that facilitate specification of covariate-varying sensitivity parameters and multiple treatments regimes.  

There are also several open technical questions about our application of Tukey's factorization to observational studies.
One important consideration for Tukey's factorization is the validity of the outcome overlap condition (Condition~\ref{cond:missing support}, Section \ref{sec:tukey_causal}).  This condition says that the support of the missing potential outcomes must be a subset of the support of the observed potential outcomes.  As discussed in \citet{franks2016non}, even when the outcome overlap condition is technically satisfied, the inferred missing data distribution can be sensitive to the estimated tails of the observed data density if the distance between the observed and missing data is large.  This is particularly evident when viewed from the importance weighting perspective, since $\frac{f(T=1\-t\mid Y(t))}{f(T=t\mid Y(t))}$ increases in $Y(t)$ in regions where the missing data density is far from the observed data density.  In this case, the inferred missing data density is largely determined by parametric assumptions about the tail behavior of the observed data densities, which often have limited information in practice.
Additionally, the outcome overlap condition may become less plausible when the covariates explain most of the variance in the observed outcome $Y(t)$.
In such a case, an unobserved confounder would only need to induce a small amount of variation in $Y(t)$ to violate the outcome overlap assumption.

In the end, the methods described in this paper are quite general, can be extended to a range of models, and are easy to interpret and implement, even for complex data generating models. We therefore believe Tukey's factorization is a powerful framework for assessing sensitivity to unobserved confounders in observational causal inference.

%% file: sections/appendix.tex
\section{Theory}
\noindent \textbf{Proof of Theorem \ref{thm:copula invariance}:} $\tau^{ATE}$, $\tau^{ATT}$, $\tau^{ATC}$ and $\tau^{OR}$ are all only functions $f(Y(t)) \quad t \in \{0, 1\}$ (or $f(Y(t) \mid X)$ for conditional treatment effects).  Thus it suffices to show that $f(Y(t) \mid T)$ are independent of any copula parameters. Note that in the extrapolation factorization we model $f(Y(t) \mid T=t)$ directly and thus, this conditional expectation is independent of copula parameters by definition.  Thus it suffices to show that $f(Y(t) \mid T=(1-t))$ is independent of copula parameters.  

\begin{align*}
f(Y(t) \mid T=(1\-t)) &= \int f(Y(t), Y(1\-t) \mid T=(1\-t)) dY(1\-t) \\
&= \int f(Y(t) \mid Y(1\-t), T=(1\-t))f(Y(1\-t) \mid T=(1\-t) dY(1\-t) \\
&\propto  \int  f(Y(t)\mid T=t)\frac{f(T=(1\-t) \mid Y(t))}{f(T=t\mid Y(t))}c(F(Y(t)\mid T), F(Y(1\-t)\mid T) \mid ) \times\\ 
&\text{   } f(Y(1\-t) \mid T=(1\-t) dY(1\-t) \\
&= f(Y(t)\mid T=t) \frac{f(T=(1\-t) \mid Y(t))}{f(T=t\mid Y(t))}\times\\
&\text{   }\int c(F(Y(t)\mid T), F(Y(1\-t)\mid T) \mid T)f(Y(1\-t)\mid T=(1\-t) dY(1\-t)\\
&= f(Y(t)\mid T=t) \frac{f(T=(1\-t) \mid Y(t))}{f(T=t\mid Y(t))}
\end{align*}
Where the last equality holds by using the definition of the copula density:
\begin{align*}
&\int c(F(Y(t)\mid T), F(Y(1\-t)\mid T) \mid T)f(Y(1\-t)\mid T=(1\-t) dY(1\-t) =\\
&= \int \frac{f(Y(t), Y(1\-t) \mid T=(1\-t))}{f(Y(t)\mid T=(1\-t))f(Y(1\-t)\mid T=(1\-t))}f(Y(1\-t)\mid T=(1\-t) dY(1\-t)\\
&= \int f(Y(1\-t) \mid Y(t), T) dY(1\-t)\\
& = 1
\end{align*}










\noindent \textbf{Proof of Proposition \ref{prop:calibration}:}  We seek to find the value of $\gamma_t$ such that the model \eqref{eqn:partial_r2}  implies $\rho^2_{Y \mid X}$ achieves a particular value, $\rho^2_*$.  
\begin{align}
    m(X, Y(t)) =: \text{logit}(e(X, Y(t))) &= \alpha_t(X) + \gamma_t Y(t))\\
    &=  \alpha_t(X) + \gamma_t\mu_t(X) + \gamma_t (Y(t) - \mu_t(X))\\
    &= \alpha_t^*(X) + \tilde \gamma_t \tilde R(t))\\
    &= m(X, \tilde R(t))
\end{align}
\noindent where $\tilde R(t) = \frac{R(t)}{\sigma_{rt}}$ is the unit-scaled complete data residual, $\tilde\gamma =: \sigma_{rt} \gamma $ and $\sigma_{rt} = \sqrt{E[Var(Y(t) \mid X)]}$.  We define $\alpha_t^*(X) =: \alpha_t(X) + \gamma_t\mu_t(X)$.  Importantly, since $m(X, Y(t))= m(X, \tilde R(t))$ the above implies that $\rho^2_{Y(t),X} = \rho^2_{R(t), X}$.  Since $\tilde R(t)$ is orthogonal to $\alpha_t^*(X)$ and has unit variance, we have $\text{Var}(m(X, \tilde R(t)) = \text{Var}(m(X) + \tilde \gamma_t \tilde R(t))) = \text{Var}(m(X)) + \tilde \gamma_t^2$.  
Thus, 
\begin{align}
\rho^2_{X, Y(t)} = \rho^2_{X, \tilde R(t)} = \frac{\text{Var}(m(X)) + \tilde \gamma_t^2}{\text{Var}(m(X)) + \tilde \gamma_t^2 + \pi^2/3.}
\end{align}
\noindent Using the definition of ``implicit R-squared'' from Section \ref{sec:calibration}, we have
\begin{align}
\rho^2_{\tilde Y(t) \mid X} &= \frac{\rho^2_{X, \tilde Y(t)} - \rho^2_{X}}
{1-\rho^2_{X}}\\
&= \frac{\frac{\pi^2/3}{\text{Var}(m(X)) + \pi^2/3} - \frac{\pi^2/3}{\text{Var}(m(X)) + \pi^2/3 + \tilde \gamma_t^2}}{\frac{\pi^2/3}{\text{Var}(m(X)) + \pi^2/3}}\\
&= 1 - \frac{\text{Var}(m(X)) + \pi^2/3}{\text{Var}(m(X)) + \pi^2/3 + \tilde \gamma_t^2}
\end{align}
Solving the above equation for $\tilde \gamma_t$ such that $\rho^2_{\tilde Y(t) \mid X} = \rho^2_*$, yields
\begin{equation}
|\tilde \gamma_t| = \sqrt{\frac{\rho^2_*}{1 - \rho^2_*}(\text{Var}(m(X)) + \pi^2/3})
\end{equation}
\noindent We complete the result by using the fact that $\tilde \gamma_t = \sigma_{rt} \gamma_t $


%% file: sections/app-figures.tex
\section{Additional Results from Section 5.1}

\begin{figure}[h]
	\centering
	\includegraphics[width=0.5\textwidth]{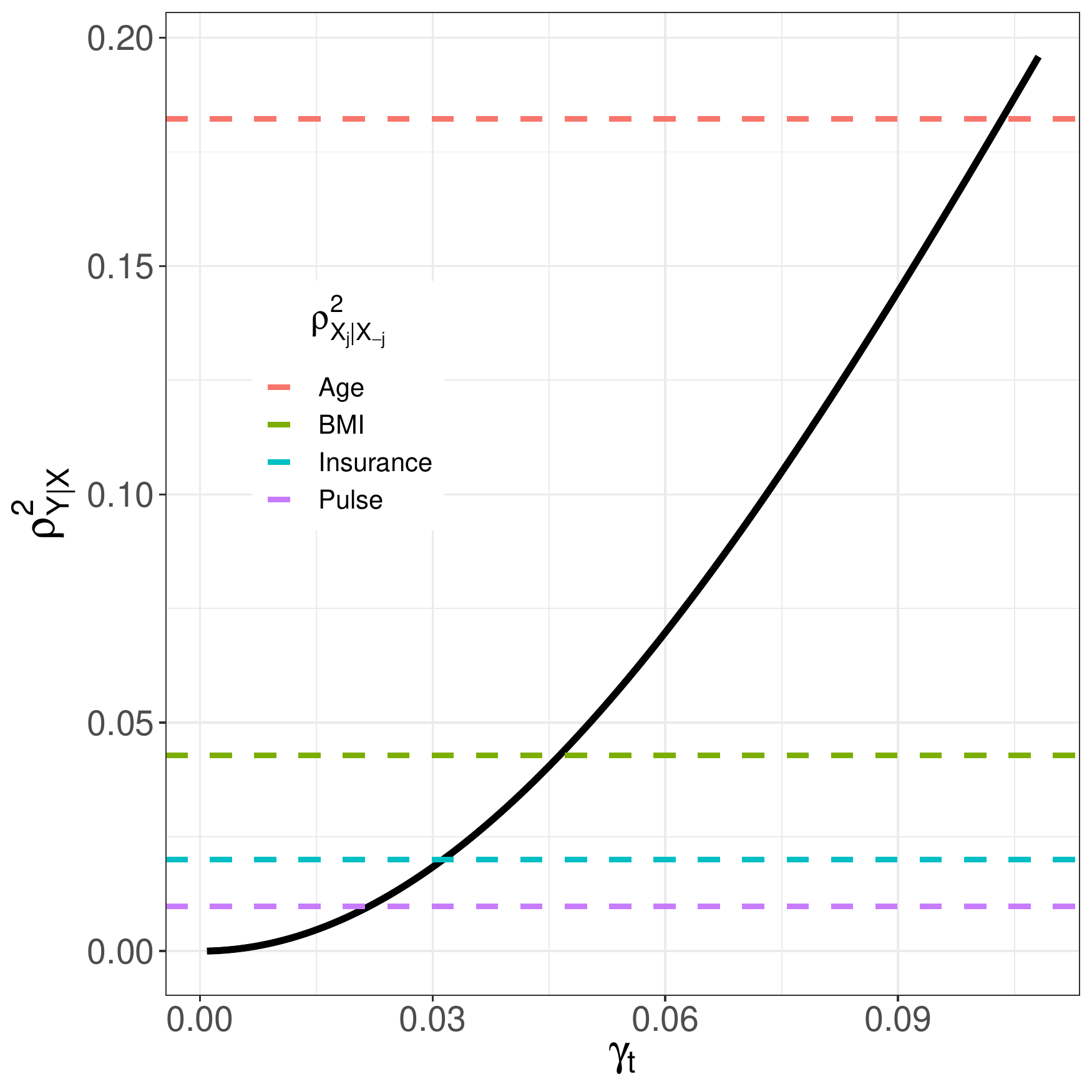}
	\caption{ $\gamma_t$ vs $\rho^2_{Y \mid X}$, calibration for the NHANES data.  The magnitude of the sensitivity parameter $\gamma_t$ is increasing with the residual coefficient of determination, $\rho^2_{Y\mid X}$. For comparison, we plot the partial coefficients of variation from covariates, $\rho^2_{X_j \mid X_{-j}}$, for the most important predictors: age, BMI, insurance and pulse.  We calibrate the magnitude of $\gamma_t$ in Section \ref{sec:analysis_nhanes} based on BMI.}\label{fig:nhanes_calibration_comparison}
\end{figure}

In Section 5.1, we focus on one particular potential outcomes model, although many plausible models are possible. In this section, we provide results for two variations of the observed potential outcome model.  This plot highlights that the ATE estimates vary as a function of both model specification (model checking) and the strength of confounding in both treatment arms (sensitivity analysis).  

\begin{figure}	

	\begin{subfigure}[t]{0.49\textwidth}
		\includegraphics[width=\textwidth]{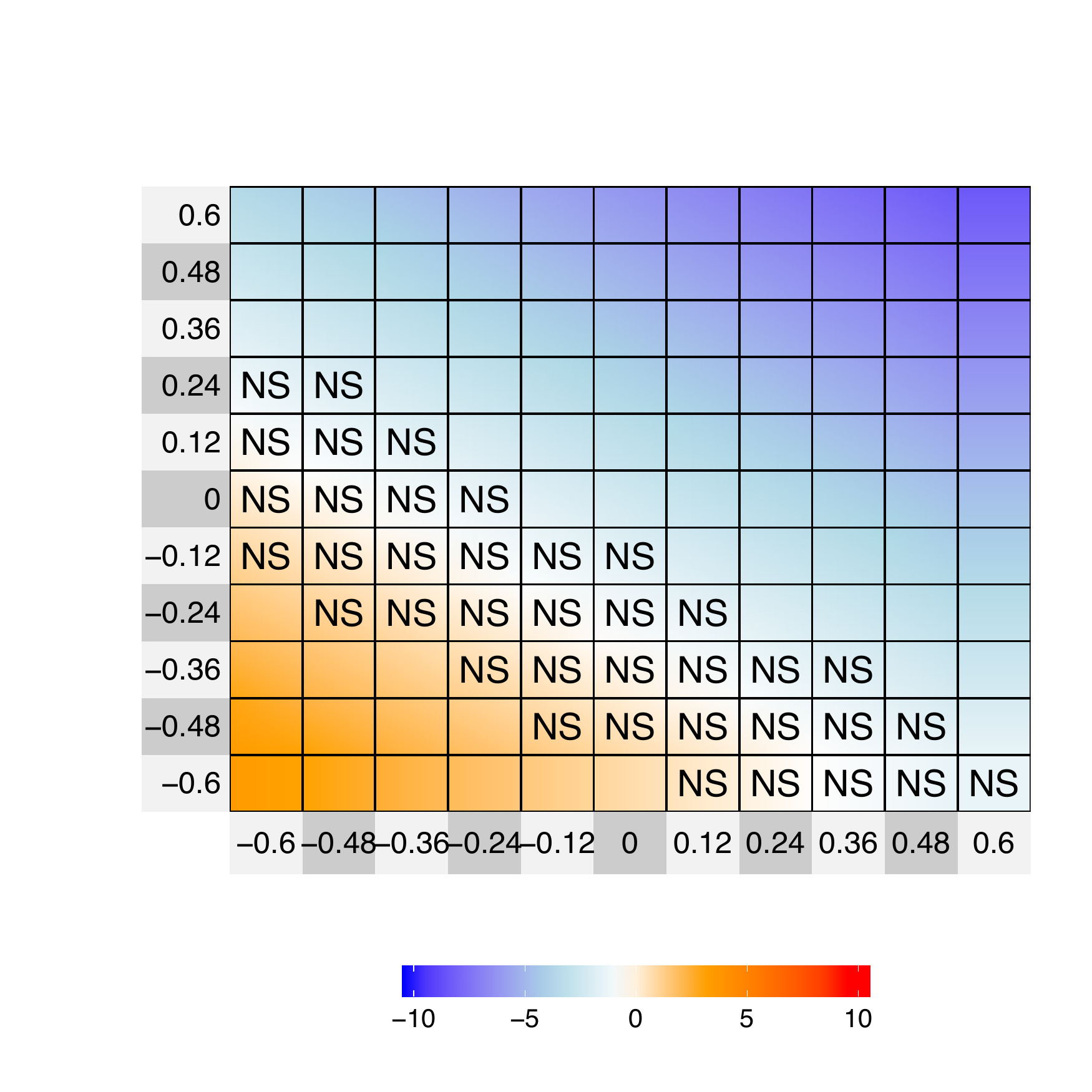}
		\caption{ATE for pooled model}\label{nhanes_ate_joint}
	\end{subfigure}
	\begin{subfigure}[t]{0.49\textwidth}
		\includegraphics[width=\textwidth]{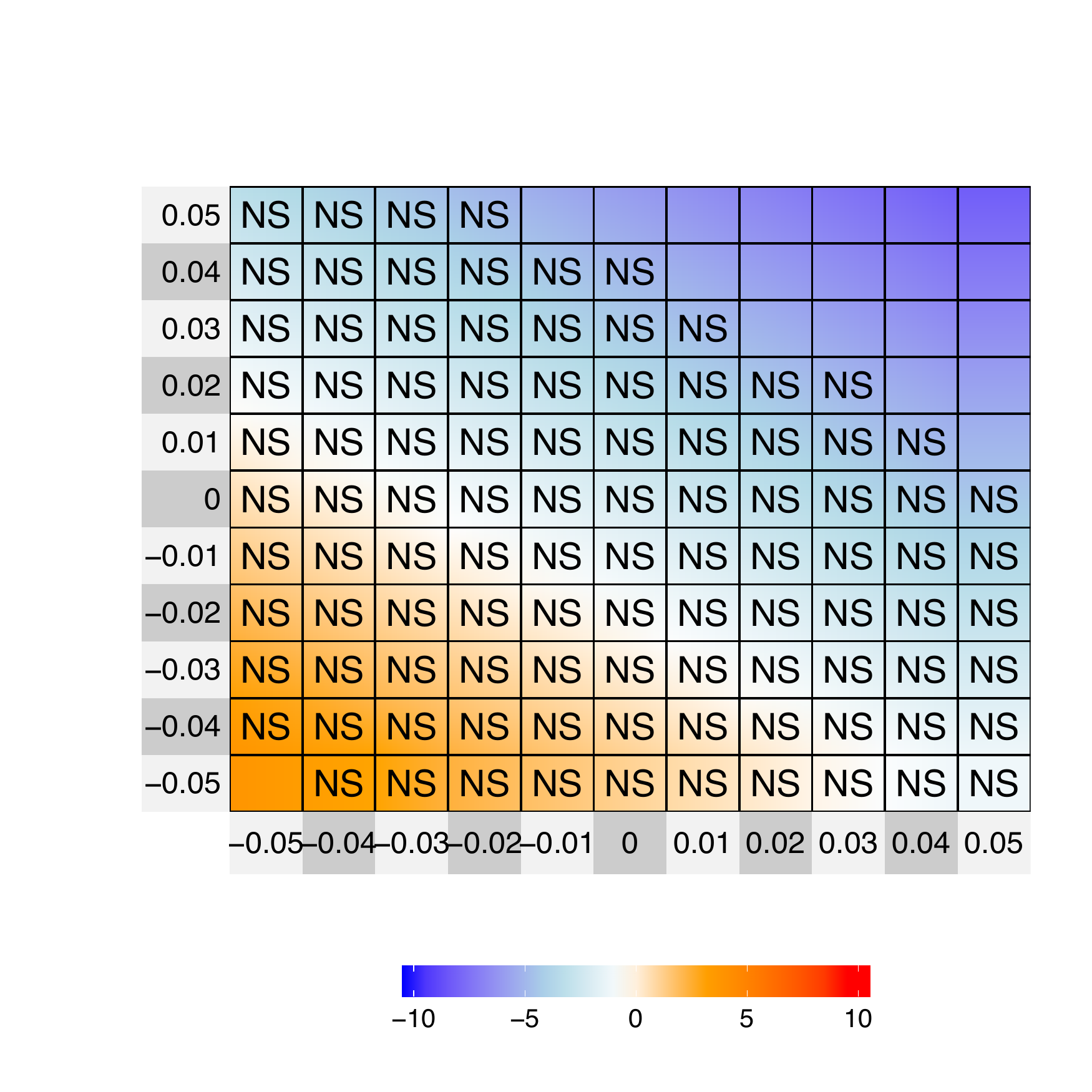}
		\caption{ATE for BCF model}\label{nhanes_ate_bcf}
	\end{subfigure}
	\caption{Average treatment effect measured in units of millimeters of mercury (mmHg). NS denotes “not significant”. a) Average treatment effect in the pooled model. Under unconfoundedness, the effect size is significantly negative.  This model has the smallest posterior uncertainty.  b) Average treatment effect in the Bayesian Causal Forest model. Although the effect sizes are comparable, the posterior uncertainty is significantly larger.}
	\label{fig:app}
\end{figure}

First, we posit a \emph{pooled} model for the mean surface and residual variance $(\mu_t(X), \sigma_t^2) \sim BART(X, T)$ with $\mu_t(X) = \mu(t, X)$ and $\sigma_t^2 = \sigma_{1\-t}^2$.  In Figure \ref{nhanes_ate_joint} we show the results for this model, which shows has the largest estimated effect size under unconfoundedness of any of the models considered.  Under unconfoundedness, the posterior mean ATE is approximately -2.5 mmHG under this model, and unlike the model proposed in Section \ref{sec:analysis_nhanes} appears significantly different from 0.  

We also show the results for the Bayesian Causal Forest (BCF) model recently introduced by \citep{hahn2017bayesian}. In this model, the observed propensity score is included as a covariate and independent BART prior distributions are specified for the control and for the heterogeneous treatment effect and one is used for the the control outcome surface.  In this model, under unconfoundedness the posterior mean for the ATE is approximately -1.73 mmHg but in contrast to the other observed data models, yields ATEs with large posterior uncertainty.  

%% file: main.bbl
\begin{thebibliography}{}

\bibitem[\protect\citeauthoryear{Abadie, Angrist, and Imbens}{Abadie
  et~al.}{2002}]{abadie2002instrumental}
Abadie, A., J.~Angrist, and G.~Imbens (2002).
\newblock Instrumental variables estimates of the effect of subsidized training
  on the quantiles of trainee earnings.
\newblock {\em Econometrica\/}~{\em 70\/}(1), 91--117.

\bibitem[\protect\citeauthoryear{Athey and Wager}{Athey and
  Wager}{2017}]{athey2017}
Athey, S. and S.~Wager (2017).
\newblock Efficient policy learning.
\newblock {\em arXiv preprint arXiv:1702.02896\/}.

\bibitem[\protect\citeauthoryear{Birmingham, Rotnitzky, and
  Fitzmaurice}{Birmingham et~al.}{2003}]{Birmingham2003}
Birmingham, J., A.~Rotnitzky, and G.~M. Fitzmaurice (2003).
\newblock {Pattern-mixture and selection models for analysing longitudinal data
  with monotone missing patterns}.
\newblock {\em Journal of the Royal Statistical Society. Series B: Statistical
  Methodology\/}~{\em 65}, 275--297.

\bibitem[\protect\citeauthoryear{Bitler, Gelbach, and Hoynes}{Bitler
  et~al.}{2006}]{bitler2006mean}
Bitler, M.~P., J.~B. Gelbach, and H.~W. Hoynes (2006).
\newblock What mean impacts miss: Distributional effects of welfare reform
  experiments.
\newblock {\em American Economic Review\/}~{\em 96\/}(4), 988--1012.

\bibitem[\protect\citeauthoryear{Blackwell}{Blackwell}{2014}]{blackwell2014selection}
Blackwell, M. (2014).
\newblock A selection bias approach to sensitivity analysis for causal effects.
\newblock {\em Political Analysis\/}~{\em 22\/}(2), 169--182.

\bibitem[\protect\citeauthoryear{Carnegie, Harada, and Hill}{Carnegie
  et~al.}{2016}]{carnegie2016assessing}
Carnegie, N.~B., M.~Harada, and J.~L. Hill (2016).
\newblock Assessing sensitivity to unmeasured confounding using a simulated
  potential confounder.
\newblock {\em Journal of Research on Educational Effectiveness\/}~{\em
  9\/}(3), 395--420.

\bibitem[\protect\citeauthoryear{{Centers for Disease Control and Prevention
  (CDC)}}{{Centers for Disease Control and Prevention (CDC)}}{1997}]{NHANES3}
{Centers for Disease Control and Prevention (CDC)} (1997).
\newblock {National Health and Nutrition Examination Survey Data III, U.S.
  Department of Health and Human Services, Centers for Disease Control and
  Prevention, Hyattsville, MD}.
\newblock \url{https://wwwn.cdc.gov/nchs/nhanes/nhanes3/DataFiles.aspx}.

\bibitem[\protect\citeauthoryear{Chernozhukov, Chetverikov, Demirer, Duflo,
  Hansen, and Newey}{Chernozhukov et~al.}{2016}]{chernozhukov2016double}
Chernozhukov, V., D.~Chetverikov, M.~Demirer, E.~Duflo, C.~Hansen, and W.~K.
  Newey (2016).
\newblock Double machine learning for treatment and causal parameters.
\newblock Technical report, cemmap working paper, Centre for Microdata Methods
  and Practice.

\bibitem[\protect\citeauthoryear{Chipman, George, McCulloch, et~al.}{Chipman
  et~al.}{2010}]{chipman2010bart}
Chipman, H.~A., E.~I. George, R.~E. McCulloch, et~al. (2010).
\newblock Bart: Bayesian additive regression trees.
\newblock {\em The Annals of Applied Statistics\/}~{\em 4\/}(1), 266--298.

\bibitem[\protect\citeauthoryear{Cinelli and Hazlett}{Cinelli and
  Hazlett}{2018}]{cinelli2018making}
Cinelli, C. and C.~Hazlett (2018).
\newblock Making sense of sensitivity: Extending omitted variable bias.

\bibitem[\protect\citeauthoryear{Cornfield, Haenszel, Hammond, Lilienfeld,
  Shimkin, and Wynder}{Cornfield et~al.}{1959}]{cornfield1959smoking}
Cornfield, J., W.~Haenszel, E.~C. Hammond, A.~M. Lilienfeld, M.~B. Shimkin, and
  E.~L. Wynder (1959).
\newblock Smoking and lung cancer: recent evidence and a discussion of some
  questions.
\newblock {\em J. Nat. Cancer Inst\/}~{\em 22}, 173--203.

\bibitem[\protect\citeauthoryear{D'Amour}{D'Amour}{2019}]{damour2019aistats}
D'Amour, A. (2019).
\newblock On multi-cause causal inference: Impossibility, sensitivity, and the
  promise of proxies.
\newblock In {\em International Conference on Artificial Intelligence and
  Statistics}, pp.\  Forthcoming.

\bibitem[\protect\citeauthoryear{Daniels and Hogan}{Daniels and
  Hogan}{2000}]{Daniels2000}
Daniels, M.~J. and J.~W. Hogan (2000).
\newblock {Reparameterizing the pattern mixture model for sensitivity analyses
  under informative dropout.}
\newblock {\em Biometrics\/}~{\em 56\/}(4), 1241--1248.

\bibitem[\protect\citeauthoryear{D{\'\i}az and van~der Laan}{D{\'\i}az and
  van~der Laan}{2013}]{diaz2013sensitivity}
D{\'\i}az, I. and M.~J. van~der Laan (2013).
\newblock Sensitivity analysis for causal inference under unmeasured
  confounding and measurement error problems.
\newblock {\em The international journal of biostatistics\/}~{\em 9\/}(2),
  149--160.

\bibitem[\protect\citeauthoryear{Ding, Feller, and Miratrix}{Ding
  et~al.}{2018}]{ding2018decomposing}
Ding, P., A.~Feller, and L.~Miratrix (2018).
\newblock Decomposing treatment effect variation.
\newblock {\em Journal of the American Statistical Association\/}, 1--14.

\bibitem[\protect\citeauthoryear{Ding and VanderWeele}{Ding and
  VanderWeele}{2016}]{ding2016sensitivity}
Ding, P. and T.~J. VanderWeele (2016).
\newblock Sensitivity analysis without assumptions.
\newblock {\em Epidemiology (Cambridge, Mass.)\/}~{\em 27\/}(3), 368.

\bibitem[\protect\citeauthoryear{Dorie, Harada, Carnegie, and Hill}{Dorie
  et~al.}{2016}]{dorie2016flexible}
Dorie, V., M.~Harada, N.~B. Carnegie, and J.~Hill (2016).
\newblock A flexible, interpretable framework for assessing sensitivity to
  unmeasured confounding.
\newblock {\em Statistics in medicine\/}~{\em 35\/}(20), 3453--3470.

\bibitem[\protect\citeauthoryear{Duan, Manning, Morris, and Newhouse}{Duan
  et~al.}{1983}]{duan1983comparison}
Duan, N., W.~G. Manning, C.~N. Morris, and J.~P. Newhouse (1983).
\newblock A comparison of alternative models for the demand for medical care.
\newblock {\em Journal of business \& economic statistics\/}~{\em 1\/}(2),
  115--126.

\bibitem[\protect\citeauthoryear{Everitt}{Everitt}{1985}]{everitt1985mixture}
Everitt, B.~S. (1985).
\newblock {\em Mixture Distributions—I}.
\newblock Wiley Online Library.

\bibitem[\protect\citeauthoryear{Franks, Airoldi, and Rubin}{Franks
  et~al.}{2016}]{franks2016non}
Franks, A.~M., E.~M. Airoldi, and D.~B. Rubin (2016).
\newblock Non-standard conditionally specified models for non-ignorable missing
  data.
\newblock {\em arXiv preprint arXiv:1603.06045\/}.

\bibitem[\protect\citeauthoryear{Gelman, Stern, Carlin, Dunson, Vehtari, and
  Rubin}{Gelman et~al.}{2013}]{gelman2013bayesian}
Gelman, A., H.~S. Stern, J.~B. Carlin, D.~B. Dunson, A.~Vehtari, and D.~B.
  Rubin (2013).
\newblock {\em Bayesian data analysis}.
\newblock Chapman and Hall/CRC.

\bibitem[\protect\citeauthoryear{Gustafson, McCandless, et~al.}{Gustafson
  et~al.}{2018}]{gustafson2018sensitivity}
Gustafson, P., L.~C. McCandless, et~al. (2018).
\newblock When is a sensitivity parameter exactly that?
\newblock {\em Statistical Science\/}~{\em 33\/}(1), 86--95.

\bibitem[\protect\citeauthoryear{Hahn, Murray, and Carvalho}{Hahn
  et~al.}{2017}]{hahn2017bayesian}
Hahn, P.~R., J.~S. Murray, and C.~M. Carvalho (2017).
\newblock Bayesian regression tree models for causal inference: regularization,
  confounding, and heterogeneous effects.

\bibitem[\protect\citeauthoryear{Hahn, Murray, and Manolopoulou}{Hahn
  et~al.}{2016}]{hahn2016bayesian}
Hahn, P.~R., J.~S. Murray, and I.~Manolopoulou (2016).
\newblock A bayesian partial identification approach to inferring the
  prevalence of accounting misconduct.
\newblock {\em Journal of the American Statistical Association\/}~{\em
  111\/}(513), 14--26.

\bibitem[\protect\citeauthoryear{Heckman}{Heckman}{1979}]{heckman1979}
Heckman, J.~J. (1979).
\newblock Sample selection bias as a specification error.
\newblock {\em Econometrica\/}~{\em 47\/}(1), 153--161.

\bibitem[\protect\citeauthoryear{Heckman, Smith, and Clements}{Heckman
  et~al.}{1997}]{heckman1997making}
Heckman, J.~J., J.~Smith, and N.~Clements (1997).
\newblock Making the most out of programme evaluations and social experiments:
  Accounting for heterogeneity in programme impacts.
\newblock {\em The Review of Economic Studies\/}~{\em 64\/}(4), 487--535.

\bibitem[\protect\citeauthoryear{Hill}{Hill}{2012}]{hill2012}
Hill, J.~L. (2012).
\newblock Bayesian nonparametric modeling for causal inference.
\newblock {\em Journal of Computational and Graphical Statistics\/}.

\bibitem[\protect\citeauthoryear{Holland}{Holland}{1986}]{holland-notes}
Holland, P. (1986).
\newblock Discussion 4: Mixture modeling versus selection modeling with
  nonignorable nonresponse.
\newblock In H.~Wainer (Ed.), {\em Drawing inferences from self-selected
  samples}, pp.\  143--149. Routledge.

\bibitem[\protect\citeauthoryear{Imai and Van~Dyk}{Imai and
  Van~Dyk}{2004}]{imai2004causal}
Imai, K. and D.~A. Van~Dyk (2004).
\newblock Causal inference with general treatment regimes: Generalizing the
  propensity score.
\newblock {\em Journal of the American Statistical Association\/}~{\em
  99\/}(467), 854--866.

\bibitem[\protect\citeauthoryear{Imbens}{Imbens}{2003}]{imbens2003sensitivity}
Imbens, G.~W. (2003).
\newblock Sensitivity to exogeneity assumptions in program evaluation.
\newblock {\em American Economic Review\/}~{\em 93\/}(2), 126--132.

\bibitem[\protect\citeauthoryear{Javaras and Van~Dyk}{Javaras and
  Van~Dyk}{2003}]{javaras2003multiple}
Javaras, K.~N. and D.~A. Van~Dyk (2003).
\newblock Multiple imputation for incomplete data with semicontinuous
  variables.
\newblock {\em Journal of the American Statistical Association\/}~{\em
  98\/}(463), 703--715.

\bibitem[\protect\citeauthoryear{Jung, Shroff, Feller, and Goel}{Jung
  et~al.}{2018}]{jung2018algorithmic}
Jung, J., R.~Shroff, A.~Feller, and S.~Goel (2018).
\newblock Algorithmic decision making in the presence of unmeasured
  confounding.
\newblock {\em arXiv preprint arXiv:1805.01868\/}.

\bibitem[\protect\citeauthoryear{Kenward}{Kenward}{1998}]{Kenward_Selection_1998}
Kenward, M.~G. (1998).
\newblock Selection models for repeated measurements with non-random dropout:
  an illustration of sensitivity.
\newblock {\em Statist. Med.\/}.

\bibitem[\protect\citeauthoryear{Klausch, van~de Ven, van~de Brug, Brakenhoff,
  van~de Wiel, and Berkhof}{Klausch et~al.}{2018}]{klausch2018estimating}
Klausch, T., P.~van~de Ven, T.~van~de Brug, R.~H. Brakenhoff, M.~A. van~de
  Wiel, and J.~Berkhof (2018).
\newblock Estimating bayesian optimal treatment regimes for dichotomous
  outcomes using observational data.
\newblock {\em arXiv preprint arXiv:1809.06679\/}.

\bibitem[\protect\citeauthoryear{Linero}{Linero}{2017}]{linero2017biometrika}
Linero, A.~R. (2017).
\newblock Bayesian nonparametric analysis of longitudinal studies in the
  presence of informative missingness.
\newblock {\em Biometrika\/}~{\em 104\/}(2), 327--341.

\bibitem[\protect\citeauthoryear{Linero and Daniels}{Linero and
  Daniels}{2015}]{linero2015flexible}
Linero, A.~R. and M.~J. Daniels (2015).
\newblock A flexible bayesian approach to monotone missing data in longitudinal
  studies with nonignorable missingness with application to an acute
  schizophrenia clinical trial.
\newblock {\em Journal of the American Statistical Association\/}~{\em
  110\/}(509), 45--55.

\bibitem[\protect\citeauthoryear{Linero and Daniels}{Linero and
  Daniels}{2017}]{linero2017bayesian}
Linero, A.~R. and M.~J. Daniels (2017).
\newblock Bayesian approaches for missing not at random outcome data: The role
  of identifying restrictions.

\bibitem[\protect\citeauthoryear{Little and Rubin}{Little and
  Rubin}{2015}]{littlebook}
Little, R.~J. and D.~B. Rubin (2015).
\newblock {\em Statistical analysis with missing data}.
\newblock John Wiley \& Sons.

\bibitem[\protect\citeauthoryear{Middleton, Scott, Diakow, and Hill}{Middleton
  et~al.}{2016}]{middleton2016bias}
Middleton, J.~A., M.~A. Scott, R.~Diakow, and J.~L. Hill (2016).
\newblock Bias amplification and bias unmasking.
\newblock {\em Political Analysis\/}~{\em 24\/}(3), 307--323.

\bibitem[\protect\citeauthoryear{Neal}{Neal}{2000}]{neal2000markov}
Neal, R.~M. (2000).
\newblock Markov chain sampling methods for dirichlet process mixture models.
\newblock {\em Journal of computational and graphical statistics\/}~{\em
  9\/}(2), 249--265.

\bibitem[\protect\citeauthoryear{Neyman}{Neyman}{1923}]{neyman1923}
Neyman, J. ({1990 [1923]}).
\newblock On the application of probability theory to agricultural experiments.
  essay on principles. section 9.
\newblock {\em Statistical Science\/}~{\em 5\/}(4), 465--472.

\bibitem[\protect\citeauthoryear{Olsen and Schafer}{Olsen and
  Schafer}{2001}]{olsen2001two}
Olsen, M.~K. and J.~L. Schafer (2001).
\newblock A two-part random-effects model for semicontinuous longitudinal data.
\newblock {\em Journal of the American Statistical Association\/}~{\em
  96\/}(454), 730--745.

\bibitem[\protect\citeauthoryear{Oster}{Oster}{2017}]{oster2017unobservable}
Oster, E. (2017).
\newblock Unobservable selection and coefficient stability: Theory and
  evidence.
\newblock {\em Journal of Business \& Economic Statistics\/}, 1--18.

\bibitem[\protect\citeauthoryear{Riddles, Kim, and Im}{Riddles
  et~al.}{2016}]{riddles2016propensity}
Riddles, M.~K., J.~K. Kim, and J.~Im (2016).
\newblock A propensity-score-adjustment method for nonignorable nonresponse.
\newblock {\em Journal of Survey Statistics and Methodology\/}~{\em 4\/}(2),
  215--245.

\bibitem[\protect\citeauthoryear{Robins, Rotnitzky, and Scharfstein}{Robins
  et~al.}{2000}]{robins2000sensitivity}
Robins, J.~M., A.~Rotnitzky, and D.~O. Scharfstein (2000).
\newblock Sensitivity analysis for selection bias and unmeasured confounding in
  missing data and causal inference models.
\newblock In {\em Statistical models in epidemiology, the environment, and
  clinical trials}, pp.\  1--94. Springer.

\bibitem[\protect\citeauthoryear{Rosenbaum}{Rosenbaum}{2017}]{Rosenbaum_Observation_2017}
Rosenbaum, P. (2017).
\newblock {\em Observation and Experiment: An Introduction to Causal
  Inference}.
\newblock Harvard University Press.

\bibitem[\protect\citeauthoryear{Rosenbaum and Rubin}{Rosenbaum and
  Rubin}{1983}]{rosenbaum1983assessing}
Rosenbaum, P.~R. and D.~B. Rubin (1983).
\newblock Assessing sensitivity to an unobserved binary covariate in an
  observational study with binary outcome.
\newblock {\em Journal of the Royal Statistical Society. Series B
  (Methodological)\/}, 212--218.

\bibitem[\protect\citeauthoryear{Rosenbaum and Silber}{Rosenbaum and
  Silber}{2009}]{rosenbaum2009amplification}
Rosenbaum, P.~R. and J.~H. Silber (2009).
\newblock Amplification of sensitivity analysis in matched observational
  studies.
\newblock {\em Journal of the American Statistical Association\/}~{\em
  104\/}(488), 1398--1405.

\bibitem[\protect\citeauthoryear{Ross and Markwick}{Ross and
  Markwick}{2018}]{dirichletprocess}
Ross, G.~J. and D.~Markwick (2018).
\newblock {\em dirichletprocess: Build Dirichlet Process Objects for Bayesian
  Modelling}.
\newblock R package version 0.2.1.

\bibitem[\protect\citeauthoryear{Rotnitzky, Scharfstein, Su, and
  Robins}{Rotnitzky et~al.}{2001}]{rotnitzky2001}
Rotnitzky, A., D.~Scharfstein, T.-L. Su, and J.~Robins (2001).
\newblock Methods for conducting sensitivity analysis of trials with
  potentially nonignorable competing causes of censoring.
\newblock {\em Biometrics\/}~{\em 57\/}(1), 103--113.

\bibitem[\protect\citeauthoryear{Rubin}{Rubin}{1974}]{rubin1974}
Rubin, D.~B. (1974).
\newblock Estimating causal effects of treatments in randomized and
  nonrandomized studies.
\newblock {\em Journal of educational Psychology\/}~{\em 66\/}(5), 688.

\bibitem[\protect\citeauthoryear{Rubin}{Rubin}{1980}]{rubin1980comment}
Rubin, D.~B. (1980).
\newblock Comment.
\newblock {\em Journal of the American Statistical Association\/}~{\em
  75\/}(371), 591--593.

\bibitem[\protect\citeauthoryear{Rubin}{Rubin}{2003}]{rubin2003basic}
Rubin, D.~B. (2003).
\newblock Basic concepts of statistical inference for causal effects in
  experiments and observational studies.
\newblock {\em Cambridge, MA: Harvard University, Department of Statistics\/}.

\bibitem[\protect\citeauthoryear{Scharfstein, Daniels, and Robins}{Scharfstein
  et~al.}{2003}]{scharfstein2003incorporating}
Scharfstein, D.~O., M.~J. Daniels, and J.~M. Robins (2003).
\newblock Incorporating prior beliefs about selection bias into the analysis of
  randomized trials with missing outcomes.
\newblock {\em Biostatistics\/}~{\em 4\/}(4), 495--512.

\bibitem[\protect\citeauthoryear{Scharfstein, Rotnitzky, and
  Robins}{Scharfstein et~al.}{1999}]{Scharfstein1999}
Scharfstein, D.~O., A.~Rotnitzky, and J.~M. Robins (1999, December).
\newblock {Adjusting for Nonignorable Drop-Out Using Semiparametric Nonresponse
  Models}.
\newblock {\em Journal of the American Statistical Association\/}~{\em
  94\/}(448), 1096--1120.

\bibitem[\protect\citeauthoryear{Van~der Laan and Rose}{Van~der Laan and
  Rose}{2011}]{van2011targeted}
Van~der Laan, M.~J. and S.~Rose (2011).
\newblock {\em Targeted learning: causal inference for observational and
  experimental data}.
\newblock Springer Science \& Business Media.

\bibitem[\protect\citeauthoryear{Wang and Blei}{Wang and
  Blei}{2018}]{wang2018blessings}
Wang, Y. and D.~M. Blei (2018).
\newblock The blessings of multiple causes.
\newblock {\em arXiv preprint arXiv:1805.06826\/}.

\bibitem[\protect\citeauthoryear{Xu, Daniels, and Winterstein}{Xu
  et~al.}{2018}]{xu2018bayesian}
Xu, D., M.~J. Daniels, and A.~G. Winterstein (2018).
\newblock A bayesian nonparametric approach to causal inference on quantiles.
\newblock {\em Biometrics\/}.

\bibitem[\protect\citeauthoryear{Zhao, Small, and Bhattacharya}{Zhao
  et~al.}{2017}]{zhao2017sensitivity}
Zhao, Q., D.~S. Small, and B.~B. Bhattacharya (2017).
\newblock Sensitivity analysis for inverse probability weighting estimators via
  the percentile bootstrap.
\newblock {\em arXiv preprint arXiv:1711.11286\/}.

\end{thebibliography}
